# Conformity of macroscopic behavior to local properties in the catalytic ammonia synthesis and oscillatory reactions on metal surfaces

**A. R. Cholach**[*]

Boreskov Institute of Catalysis, Prospekt Akademika Lavrentieva 5, 630090 Novosibirsk, Russian Federation

## Abstract

Unique catalytic potential makes transition and noble metals the permanent objects for basic and applied research. A great number of experimental and theoretical studies enable to find out the general regularities in a field. The present research carries out this duty on the grounds of strong interrelation between catalytic, kinetic and thermodynamic behaviour of the reaction system. The trials of the catalytic ammonia synthesis and the oscillatory $NO+H_2$ reactions demonstrate that the peculiar macroscopic kinetics and catalytic activity are provided by the local thermodynamics.

Structure and activity of catalytic sites are correlated within a realistic model, where total undercoordination of adjacent surface atoms and enthalpy of local reaction is taken as a descriptor for structure and activity, respectively. In case of ammonia synthesis, the model has specified the resonant catalytic centers on metal surfaces in close agreement with relevant experimental data. The basal planes of noble metals are less active than Fe- and Ru-based catalysts, whereas an extraordinary activity of small Pt, Ir and Rh clusters can be expected. A strong advantage of imperfections compared to perfect areas in the surface wave nucleation during sustained spatiotemporal phenomena is evaluated.

Isothermal rate oscillations in open heterogeneous catalytic reaction systems are expected under the multiplicity of reaction intermediates fairly different in activity, providing the steady state and reaction rout multiplicity. Switching between active and inactive kinetic brunches gives rise to the explosive coverage changeover that can be visualized as a traveling wave. A single pattern of oscillations in the $NO+H_2$ reaction includes the key role of $NH_{ad}$ intermediates that should follow the "easy-come-easy-go" principle, providing fast catalytic removal of strongly bound nitrogen from the saturated layer. Mathematical simulations have revealed three kinetic region-attractors of the steady state, regular oscillations, and complete reaction inhibition by the oxygen atoms. The driving forces, the feedback, and chemical interactions within the traveling waves are clearly understood.

---

[*] Email: cholach@catalysis.ru

## Introduction

The concept of active sites is fundamental for heterogeneous catalysis since the most of catalytic surfaces are not uniformly active, but exhibit activity only at ensembles of special arrangement or chemical composition of near-surface atoms [1-3]. For example, the distribution of nitrogen atoms over the Ru(0001) surface after NO dissociation indicates that the active sites are formed by the topmost low coordinated metal atoms of atomic steps; however, experiments demonstrate a complex manner, in which the structure of a catalytic surface determines the activity of the catalyst, and confirm the active site concept [2].

Surface defects of different origin, chemical nature and morphology play a complex role in heterogeneous catalysis being the active sites for target reactions or inhibitor accumulation, the traps for charged species, etc. [4, 5]. Availability of defects may be the necessary condition for adsorption and different catalytic processes [6, 7]. The relevance of the structure gap between idealized models and real catalysts with various surface defects was evident for decades, but systematic studies of such surfaces have begun only after reasonable understanding of the model systems [8]. Defects can influence the reactivity via geometric or electronic features related to the specific surface structure, affect the coverage, and stabilize the activated states [9]. The number of experimental and theoretical evidence of extremely high reactivity of the certain defects on transition metal surfaces grows continuously [2, 10]. For example, the CO molecule cannot dissociate on the Rh(111) surface, while the presence of steps on this surface decreases the activation barrier by ~120 kJ/mol and enables the C-O bond breaking [10, 11]. Steps decrease the energy barriers of $N_2$ and NO dissociation by 100 kJ/mol in comparison with the most close-packed surfaces of Ru, Fe, Mo and Pd [10]. Site stabilization effect 50-80 kJ/mol at edges and corners of Pt nanoparticles against the Pt(111) plane was reported for $CH_n$ species [12]. CO oxidation and $O_2$ reduction on the Pt(111) in the alkaline media serve as a model of defect-favored and terrace-favored reaction, respectively [9]. Catalytic NO reduction by $H_2$ on transition metal surfaces has received much attention due to its importance for the automobile exhaust control [13, 14]. The study of NO + $H_2$ reaction on the Rh tip using the Field Emission Microscopy revealed that surface defects and probably grain boundaries are responsible for initiation of the chemical waves caused by the large difference in N-Rh bond strength and, hence, in reactivity [15]. It was observed that propagating wave fronts during the CO+$O_2$ reaction on the polycrystalline Pt foil are confined by grain boundaries [16], alike the case of CO pressure variations [17, 18]. The CO+$O_2$ reaction on Pt(110) and Pt(210) is actually governed by the structural defects [19], while the point and extended defects can become active sites for adsorption, surface reactions and cluster nucleation [20]. The importance of defects is indicated by frequent variation of activity with particle size. This fact has led to a conclusion about existence of structure-sensitive reactions [1, 21-22].

It turns out that the substantial part of basic concepts of heterogeneous catalysis has been discovered during development of the Haber–Bosch process, and the catalytic ammonia synthesis can be regarded as a benchmark in the field [23, 24]. In particular, the highest activity of Fe(111) and (211) single crystals among others was attributed to "$C_7$ sites" corresponding to 7-coordinated Fe atom in the second layer [22, 25]. Similar Ru sites exhibited no specific catalytic activity in the ammonia synthesis, and enhanced activities of certain Ru crystal faces were related to their comparatively large roughness and openness [21]. The subsequent papers reported responsibility of "$B_5$ sites" for high activity of Ru-based catalysts in $NH_3$ synthesis [26-28]. However, the exact nature of active sites as well as the mechanism of their catalytic promotion remains a matter of speculation [2].

Oscillatory performance of numerous chemical systems at fixed experimental conditions seems to destroy fundamentals of reaction kinetics at a first glance. Continuous studies of isothermal rate and spatiotemporal oscillations have been performed over the last three decades to clear up the matter [19, 29, 30]. Catalytic NO reduction by $H_2$ on transition metal surfaces has received much attention due to its importance for the automobile exhaust control. [13, 14]. The astonishing phenomenon of "surface explosion" in Fig. 1 followed by the sustained traveling waves over the entire surface were observed by Field Emission Microscopy (FEM) in the course of the $NO+H_2$ reaction on Rh tip, where the wave nucleation always proceeds at surface defects like the grain boundary [31].

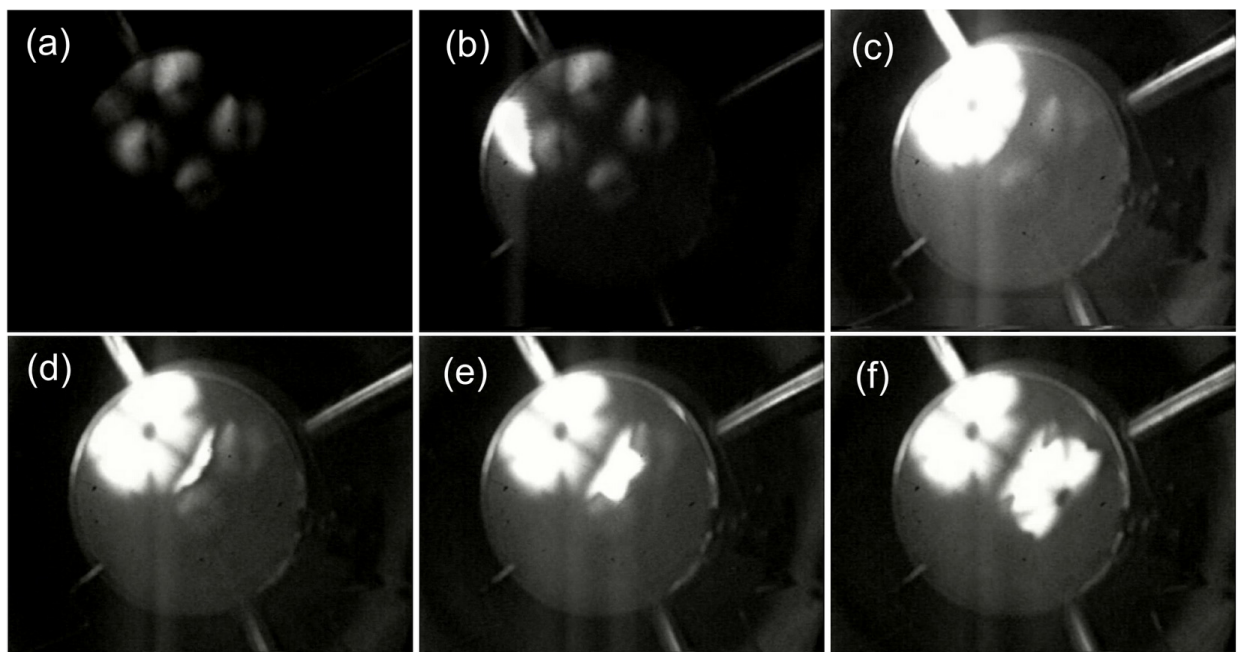

**Fig. 1** FEM screenshots during the "surface explosion" on a Rh twin tip in the mixture of $P_{NO}$ = $1.1 \cdot 10^{-5}$ and $P_{H2}$ = $1.3 \cdot 10^{-4}$ Pa at T= 464 K (a) 0 s, (b) 0.1 s, (c) 0.5 s, (d) 1.0 s, (e) 2.0 s (f) 3.5 s [32]. Regular (a-f) pictures rotation is further observed at fixed T = 432 K [31].

Similar spatiotemporal oscillations on the Pt and Ir tips [33, 34] and sustained isothermal rate oscillations were revealed for the NO+$H_2$ reaction on various noble metal surfaces [30, 35]. The detailed knowledge on oscillation mechanism clarifies the steady state processes that are of the current and potential importance as essential parts of the automotive exhaust catalysis [30, 36, 37].

Unique catalytic properties of the transition and noble metal surfaces encourage a great number of fundamental and applied studies. This paper is aimed to find out general regularities in the field on the basis of strong interrelation between catalytic, kinetic and thermodynamic behaviour of a reaction system. The current trials of the catalytic $NH_3$ synthesis and oscillatory NO+$H_2$ reaction will demonstrate a vivid pattern of that interrelation resulting in peculiar macroscopic kinetics and special catalytic activity. The catalytic ammonia synthesis is a conventional touchstone for novel approaches in surface science, while noble metals form the active component of current automobile three-way catalysts to control the NO emission among others.

## 1. Theoretical

Thermodynamic properties of the adsorbed species were studied by semi-empirical Method of Interacting Bonds (MIB), which considers any polyatomic system as a set of two-center bonds [38-42]. Free atoms correspond to $H_0|\Psi_0\rangle = E_0|\Psi_0\rangle$, where $H_0$ and $\Psi_0$ are the Hamiltonian and the wave function of a ground state with energy $E_0$, respectively. The formation of a single $i$-th bond between adjacent atoms can be described with the perturbed potential $W_i^{'}$, the wave function disturbance $\varphi_i$ and the energy change $\varepsilon_i^{'}$ – by $(H_0 + W_i^{'})|(\Psi_0 + \varphi_i)\rangle = (E_0 - \varepsilon_i^{'})|(\Psi_0 + \varphi_i)\rangle$. Setting $\langle\varphi_i|\Psi_0\rangle = 0$ and $\langle\Psi_0|\Psi_0\rangle = 1$ do not affect generality, then $W_i^{'} = -\varepsilon_i^{'} + \dfrac{(E_0 - H_0)|\varphi_i\rangle}{\Psi_0 + \varphi_i} = -\varepsilon_i^{'} + W_i$. The arbitrary number of bonds similarly corresponds to $H = H_0 + \sum_i W_i^{'} = H_0 - \sum_i \varepsilon_i^{'} + \sum_i W_i$ with wave function of $\Psi = \Psi_0 + \sum_i \nu_i \varphi_i$ superposition. All $\varphi_i$ and, therefore, $W_i$ are assumed non-zero around $i$-th bond only, and a $\nu_i$ set corresponds to the minimal energy $E = \dfrac{\langle\Psi|H|\Psi\rangle}{\langle\Psi|\Psi\rangle}$. Relevant calculation result in basic relations for atomization enthalpy [40]:

$$H_{at} = \sum_i \nu_i(2 - \nu_i)E_i - \sum_{i>k}\sum \nu_i\nu_k\Delta_k \; ; \; \frac{dH_{at}}{d\nu_i} = 0 \; , \qquad (1)$$

where $E_i \equiv \dfrac{\langle\Psi_0|W_i|\Psi_0\rangle}{1 + \sum_i\sum_k \nu_i\nu_k\langle\varphi_i|\varphi_k\rangle}$, $\Delta_k \equiv \dfrac{2\langle\varphi_i|H_0 - E_0 + W_i + W_k|\varphi_k\rangle}{1 + \sum_i\sum_k \nu_i\nu_k\langle\varphi_i|\varphi_k\rangle}$ and the most $\varepsilon_i^{'} << E_i$.

The bond coefficients $0 < \nu_i < 1$ follow from $i$ linear equations corresponding to $H_{at}$ maxima. Empirical parameters $E_i$, $\Delta_{ik}$ are found from Eq. (1) and the reference $H_{at}$; $E_i$ and $\Delta_{ik}$ are set invariable for $i$-th bond and for atom, where $i$-th bond meets $k$-th one, respectively. The energy of the $i$-th bond breakage is $\sim\nu_i E_i$, and $E_i$ may be regarded as bond energy of the free two-atomic particle ($\nu_i = 1$). The exact bond energy accounts for removed repulsions and equals to difference between $H_{at}$ values of a system with and without $i$-th bond. The larger is $\nu_i$ the stronger is $i$-th bond, while negative $\nu_i$ means thermodynamic prohibition for a given bond formation.

MIB does not predict molecular or crystal structures, atomic charges and other particular features, but reveals perfect comparative accuracy and a great time saving in comparison with the conventional Density Functional Theory (DFT). Equality of $E_i$, $\Delta_{ik}$ parameters of the reference and examined molecules (i.e. the same chemical bond nature) is the only condition for high reliability of MIB calculations demonstrated in Table 1 [38].

**Table 1** Experimental [43] and calculated atomization enthalpies $\Delta H_{at}$ of hydrocarbons (kJ/mol)

| Molecule | $\Delta H_{at}$ (exp.) | $\Delta H_{at}$ (calc.) | Difference (%) |
|---|---|---|---|
| *$CH_4$ | 1663.3 | 1654.9 | - 0.5 |
| *$C_2H_6$ | 2825.4 | 2831.6 | + 0.2 |
| $C_3H_8$ | 3997.9 | 3999.7 | + 0.0 |
| n-$C_4H_{10}$ | 5172.5 | 5168.7 | - 0.1 |
| n-$C_5H_{12}$ | 6346.4 | 6337.6 | - 0.1 |
| n-$C_6H_{14}$ | 7519.1 | 7506.5 | - 0.2 |
| n-$C_7H_{16}$ | 8692.5 | 8675.4 | - 0.2 |
| *c-$C_6H_{12}$ | 7039.6 | 7013.4 | - 0.4 |

*$\Delta H_{at}$ of these molecules were used to determine operating parameters $E_{CC}$ = 569.4; $E_{CH}$ = 565.4 and $\Delta_C$ = 138.2 kJ/mol.

Operating $E_{MM}$ and $\Delta_M$ parameters in Table 2 were determined by Eq. (1) using reference enthalpies of bulk phases formation and free $NH_3$ molecule; the empirical correlation $E_{MM}/\Delta_M$ = 4.3 was used for each substrate [32, 38]. The heats of $H_2$ adsorption are taken from reference experimental data and set independent on the coverage and substrate structure [39, 44]. Of course, $Q_{H2}$ can be affected by lateral interaction, but the $\underline{H}$ coverage under relevant experimental conditions is too low to be of importance. Accounting for $Q_{H2}$ dependence on structure also seems unreasonable since it would just complicate the content and slightly change the calculated data but not affect the conclusions.

**Table 2** Operational parameters and heats of $H_2$ adsorption $Q_{H2}$ (kJ/mol)

| | $Q_{H2}$ | $E_{MM}$ | $\Delta_M$ | $E_{MN}$ | $E_{NH}$ | $\Delta_N$ |
|---|---|---|---|---|---|---|
| Pt | 67.0 | 334.3 | 77.7 | 384.9 | 596.4 | 313.8 |
| Rh | 77.8 | 333.0 | 77.4 | 405.8 | | |
| Ir | 83.2 | 392.9 | 91.4 | 453.1 | | |
| Ru | 87.0 | 381.1 | 88.6 | 520.4 | | |
| Re | 83.4 | 456.6 | 106.2 | 555.3 | | |
| Fe bcc | 83.0 | 273.5 | 63.6 | 404.5 | | |
| Fe fcc | 83.0 | 246.3 | 57.3 | 404.5 | | |

Parameters $E_{PtN}$ and $E_{RhN}$ in Table 2 were determined indirectly, because Pt and Rh do not form bulk nitrides [45, 46]. These parameters were verified by comparison of calculated and experimental activation energy for $\underline{N}$ desorption. Diffusion proceeds via different bonding of the adsorbed species to the surface as represented in Fig. 2. The activation energy $E_{diff}(\underline{N})$ was estimated as a difference between the formation enthalpy of two adjacent $\underline{N}$ states with different number of bonds with the surface atoms [45, 46]; then the desorption energy is a double activation barrier $E_{diff}(\underline{N})$, assuming the mobility of both $\underline{N}$ atoms is necessary under combination. Similar approach has enabled proper description of experimental self-diffusion activation energies for the transition metal surfaces [47]. Table 3 testifies to reliability of parameters $E_{PtN}$, $E_{RhN}$ by the admissible correspondence between the calculated and experimental data.

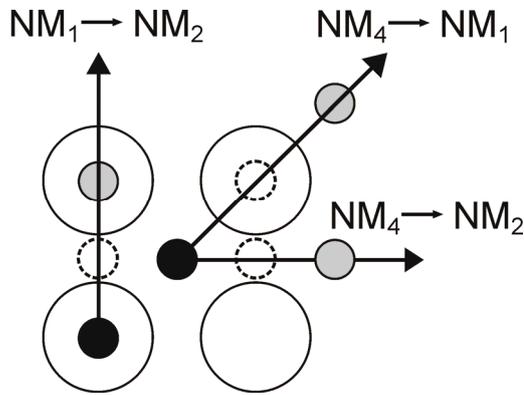

**Fig. 2** Diffusion of the adsorbed particles (small circles) between equivalent initial (black) and final (grey) sites on the fcc (100) plane includes intermediate states (dashed) with different number of bonds to surface atoms (large circles).

**Table 3** Calculated and experimental activation energy for $\underline{N}$ desorption (kJ/mol)

| Plane | Calculated* | Experimental |
|---|---|---|
| Pt(111) | 72.0-82.8 | 83.4 [48, 49] |
| Rh(100) | 142.0-167.2 | 164.7 [50] |
| Rh(111) | 83.6-100.4 | 108.7-125.4 [51] |
| Rh(110)-(1x1) | 121.2 | 115.0 [52] |
| Rh(110)-(2x1) | 117.8 | 117.9 [52] |

*Smaller and larger values correspond to nitrogen coverage $\theta_N = 1$ and 0 ML, respectively; $\theta_N = 0.5$ ML is considered for the Rh(110) plane.

## 2. Structure-activity interrelation for catalytic centers

Specific behaviour of surface imperfections in heterogeneous catalysis is a regular matter for qualitative consideration. According to the general rule, the step sites exhibit enhanced catalytic activity, if a metal surface is not highly active and vice versa, the role of undercoordinated sites is much less essential in case of highly reactive metals [8]. The Taylor' "Theory of the Catalytic Surface" emphasizes the important role of defects formed by atoms with "unsaturated valences" [53]. This concept has been successfully developed by means of experimental and theoretical studies [2, 3, 7]. It was specifically suggested that the *total* valence unsaturation (in comparison with bulk atoms) can be a realistic and effective measure of imperfection of the adsorption site [38]. Importance of this topic and the lack of quantitative data have encouraged a systematic study of surface defects on basal planes of transition and noble metals with respect to the $N_2$ dissociative adsorption and $N_{ad} + H_{ad} \rightarrow NH_{ad}$ hydrogenation. The main purpose was to clarify the role of extended defects and to specify the optimal catalytic centers in relevant reactions of $NO+H_2$ and $NH_3$ synthesis, respectively. The catalytic activity is contributed by a great number of independent and correlated parameters, and importance of each one depends on particular conditions. The present study focuses solely on imperfections with no intention to depreciate the significance of perfect areas.

### 2.1. Structural model

Accept a set of $n$ adjacent atoms $M$, each bound to the adsorbed species, as an adsorption site or catalytic center $M_n$. Each atom at the perfect planes in Fig. 3 is characterized by the number of lost bonds $\eta$ in comparison with the bulk atom. The sum of bonds $\Sigma = \sum_1^n \eta_i$ lost by atoms in $M_n$ site is taken for evaluation of the local surface imperfection, while the heat of particular reaction or adsorption at that site is considered as a measure of activity in catalytic promotion or adsorption, respectively. The latter follows from Brønsted-Evans-Polanyi (BEP) relation: the higher is the heat of reaction, the lower is the activation barrier and, therefore, the more active the process. Similar correlation is widely used elsewhere [54-56].

According to Fig. 3, the monoatomic site $M_1$ at fcc (111) plane has $\Sigma = 3$, the site $M_3$ consisting of 3 atoms has $\Sigma = 9$. Similarly, the fcc (110)-(1×1) plane can provide $\Sigma = 1$ and 5 for sites $M_1$; $\Sigma = 2$, 6 and 10 for sites $M_2$; $\Sigma = 11$ for site $M_3$ and $\Sigma = 12$ for site $M_4$, etc.

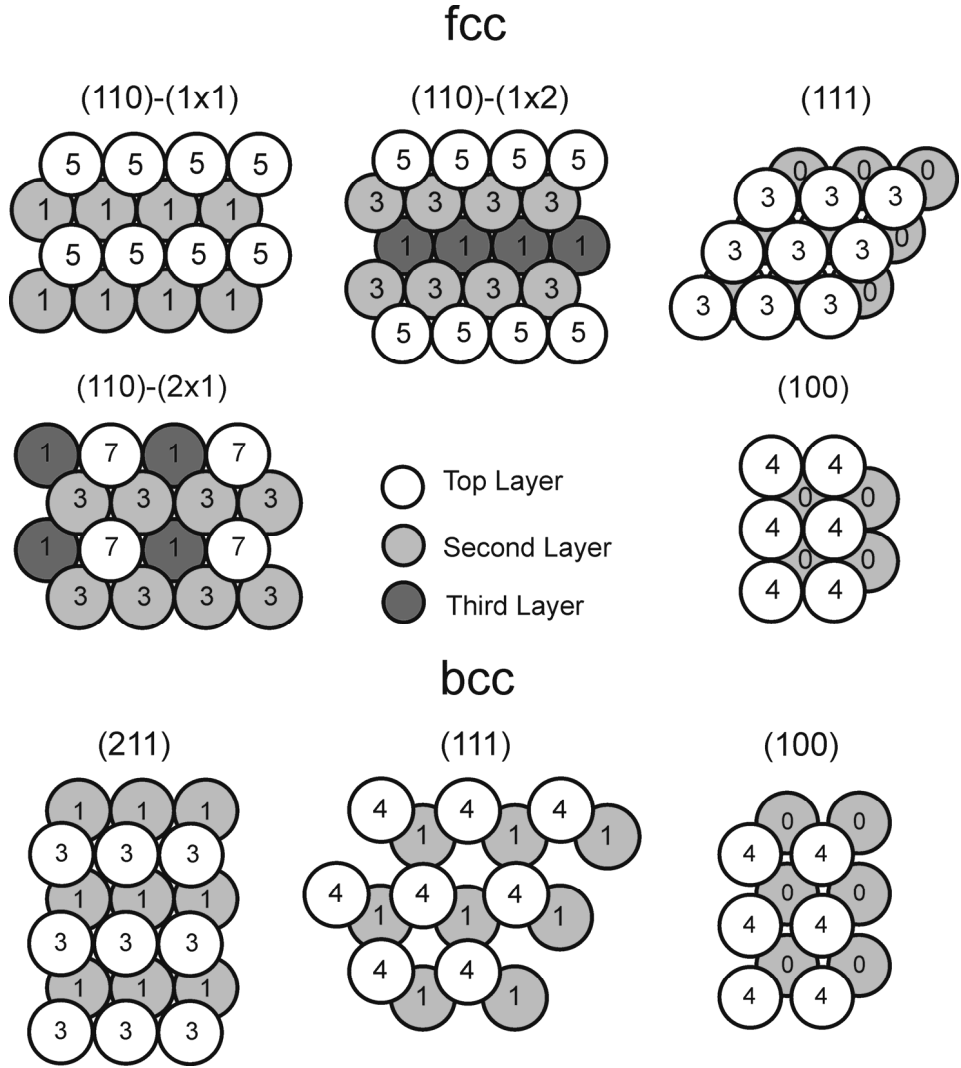

**Fig. 3** Individual numbers of bonds $\eta$ lost by each atom on basal planes are indicated.

Thermodynamic properties of the following processes have been studied by MIB:

$$\frac{1}{2}\,H_2 \quad\longleftrightarrow\quad \underline{H} + \frac{1}{2}\,Q_{H2}$$

$$\frac{1}{2}\,N_2 + \frac{1}{2}\,H_2 + M_n \quad\longrightarrow\quad \underline{NH}(M_n) + Q_{NH} \qquad (2)$$

$$\frac{1}{2}\,N_2 + M_n \quad\longleftrightarrow\quad \underline{N}(M_n) + Q_N \qquad (3)$$

$$\underline{N} + \underline{H} \quad\longrightarrow\quad \underline{NH} + Q_{\underline{NH}} \qquad (4)$$

$$2\,\underline{NH} \quad\longrightarrow\quad N_2 + H_2 + Q_{2\underline{NH}} \qquad (5)$$

$$\underline{N} + \underline{NH} \quad\longrightarrow\quad N_2 + \underline{H} + Q_{\underline{NNH}} \qquad (6)$$

Hereinafter the adsorbed states are underlined in Italic; $M_n$ is the adsorption/catalytic site consisting of $n$ adjacent atoms; H atom in $\underline{NH}$ particle is bound to N and not to M atom; $H_2$ adsorption is assumed to be equilibrium; both equilibrium and dynamic conditions for $\underline{N}$ and $\underline{NH}$ species are considered.

Atomization enthalpy $H_{at}$ of $\underline{N}$ and $\underline{NH}$ species at the given $M_n$ site is a direct result of calculation (1). Free atoms at standard conditions (298.15 K; 1 at) are taken as a reference state in left and right sides of the reactions (2-4); then heats of those reactions $Q_{NH}$, $Q_N$ and $Q_{\underline{NH}}$ have been found as follows:

$$Q_{NH} = H_{at}(\underline{NH}) - H_{at}(M_n) - \tfrac{1}{2} H_{at}(N_2) - \tfrac{1}{2} H_{at}(H_2)$$

$$Q_N = H_{at}(\underline{N}) - H_{at}(M_n) - \tfrac{1}{2} H_{at}(N_2)$$

$$Q_{\underline{NH}} = H_{at}(\underline{NH}) - H_{at}(\underline{N}) - \tfrac{1}{2} Q_{H2}$$

The calculations were performed for about 300 $M_n$ sites ($M$ = Pt, Rh, Ir, Re, Fe and Ru; n = 1; 2; 3; 4; 5) at perfect (111), (100), (110)-(1×1), (110)-(1×2), (110)-(2×1) planes of the face centered cubic (fcc) structure and at (111), (100), (211) Fe planes of the bulk centered cubic (bcc) structure. The bond parameters of the adsorbed species as well as $Q_N$ (3) and $Q_{\underline{NH}}$ (4) values gradually change by ~10% with the coverage increase from $\theta$ = 0 to 1 monolayer (ML) [39]. Figure 4 gives an example related to the bond strengths (i.e. bond coefficients) in $\underline{NH}$ species. The further results concern the uniform adsorbed layers with $\theta_N = \theta_{NH} = 0.5$ ML with respect to the unreconstructed surface.

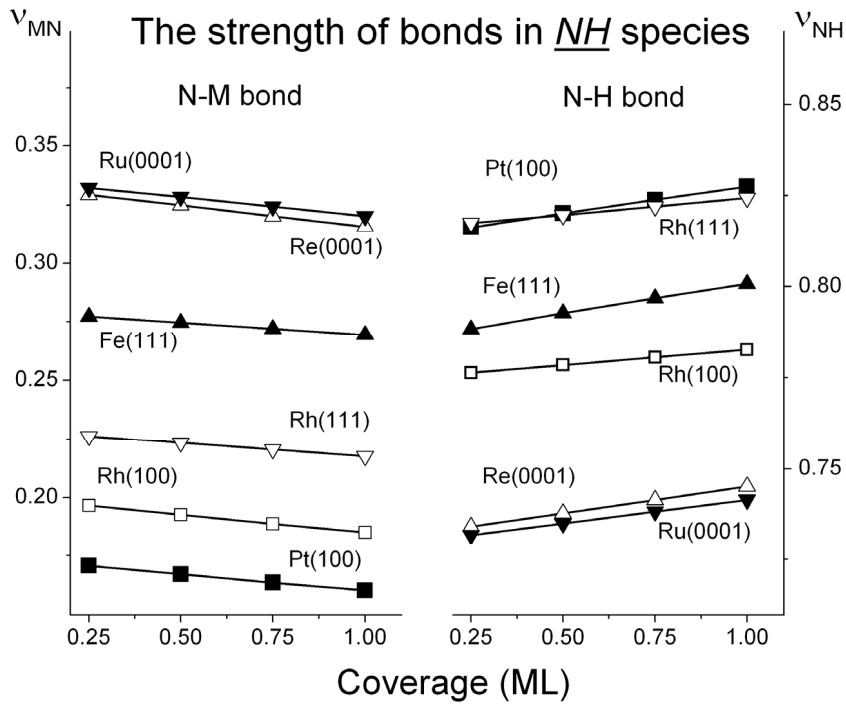

**Fig. 4** Bond coefficients (bond strengths) $\nu_i$ for the most strongly bound $\underline{NH}$ species on fcc planes as functions of $\underline{NH}$ coverage.

The repulsion between M-N bonds at the atom $M$ occurs at $M_4$ site on fcc (100) plane, at $M_4$, $M_3$, $M_2$ sites on fcc (110)-(2×1) plane, etc., and does not occur at $M_2$, $M_3$ sites on fcc (110)-(1×1) plane in Fig. 3, etc. Particular consideration reveals the repulsion energy $m\nu_{MN}\dfrac{E_{MN}\Delta_M}{2E_{MN}+\Delta_M}$, which was taken off the heats of reactions (3) and (4) to keep the quantification of different adsorption sites similar (here $m=1$; 2 is a number of extra repulsions per N-containing adsorbed particle); the single repulsion energy is 10-20 and 5-10 kJ/mol for $\underline{N}$ and $\underline{NH}$ species, respectively.

The results of calculations related to reactions (3) and (4) at 2-, 3- and 4- atomic sites demonstrate the linearity of $Q_N(\Sigma)$ and $Q_{\underline{NH}}(\Sigma)$ dependencies in Fig. 5:

$$Q_N(\Sigma) = A + \alpha\Sigma$$

$$Q_{\underline{NH}}(\Sigma) = C - \gamma\Sigma \tag{7}$$

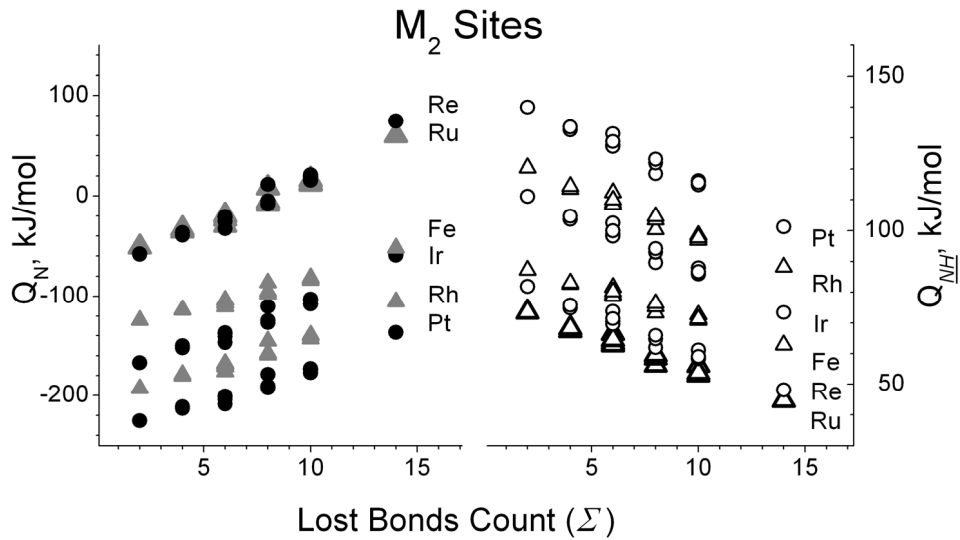

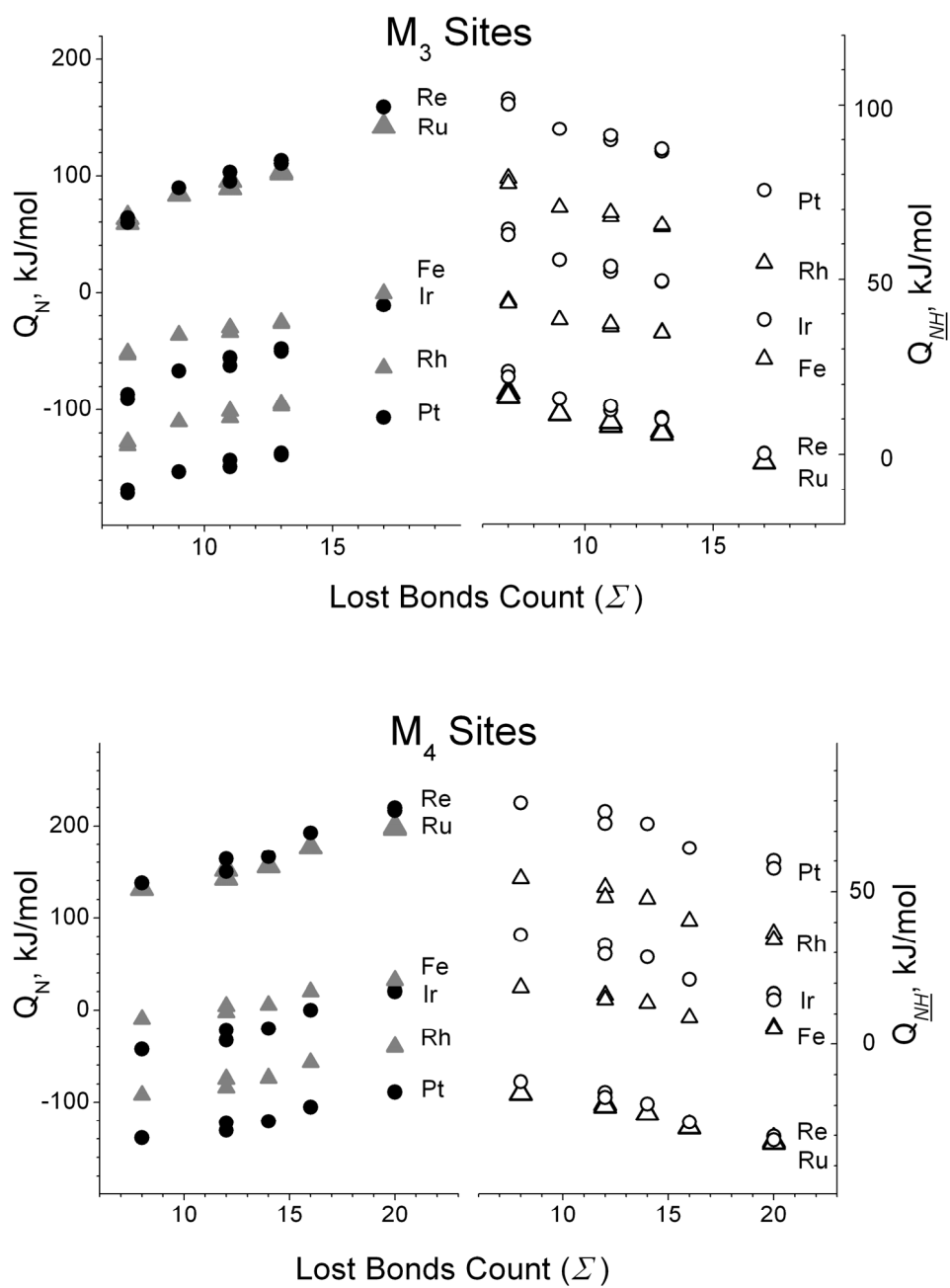

**Fig. 5** The heat of $N_2$ adsorption (reaction (3); filled symbols) and $N$ hydrogenation (reaction (4); open symbols) at 2-, 3- and 4-atomic sites on fcc planes of Pt, Ir, Re (circles) and Rh, Fe, Ru (triangles).

Parameters $A$, $\alpha$ and $\gamma$ in Table 4 needed for subsequent analysis are determined from linear regressions in Fig. 5. The opposite $Q_N(\Sigma)$ and $Q_{NH}(\Sigma)$ behavior characterizes each site. As an example, $Q_N(\Sigma)$ and $Q_{NH}(\Sigma)$ functions for different number of Ru atoms in a site are compared in Fig. 6. Other substrates with close packed structure demonstrate similar quantitative evidence of a common rule regarding optimal catalytic center [57]: the less tightly $\underline{N}$ atom is bound to surface, the more readily it undergoes further hydrogenation (4). This regularity concerns isolated molecules or ensembles as well; in particular, the supported or free $M_2$ ($\Sigma = 22$), $M_3$ ($\Sigma = 30$) and $M_4$ ($\Sigma = 27$ or 40 for pyramidal or plane structure, respectively) clusters should exhibit extraordinary activity in the $N_2$ dissociative adsorption.

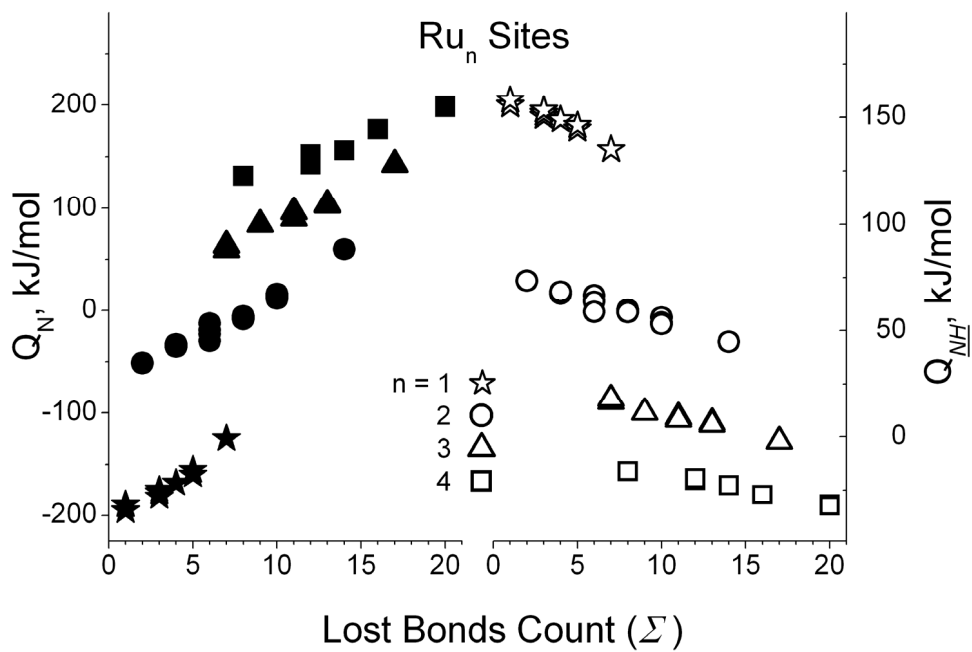

**Fig. 6** The heat of $N_2$ adsorption (reaction (3); filled symbols) and $\underline{N}$ hydrogenation (reaction (4); open symbols) at $n$-atomic sites of Ru fcc planes.

The larger $\Sigma$ and $Q_N$ indicate the more open undercoordinated site as the most preferable for $\underline{N}$ species (Fig. 3, 5). The low activity of (100) steps and (111) terraces in NO dissociation on Rh(533) and (311) single crystal surfaces [58] can result from difference between relative $Q_N$ values. On the contrary, deepened sites with lower $\Sigma$ provide smaller $Q_N$ (Fig. 3, 5). The particular spot (characterized by the large $\Sigma$ and weak $\underline{N}$ bonding between step and terrace in comparison with the plane surface) is preferable for $N_2$ desorption, and that is probably why it occurs at the boundary of (111) terraces and (100) steps in the Rh(533) single crystal according to spatial distribution measurements [59].

**Table 4** Parameters of linear $Q_N(\Sigma)$ and $Q_{NH}(\Sigma)$ functions (7) for $i$-atomic sites $M_i$ (kJ/mol)

| | Pt | Rh | Ir | Ru | Re | Fe fcc |
|---|---|---|---|---|---|---|
| | | | $M_2$ | | | |
| A | -244.0 | -211.0 | -189.8 | -73.6 | -85.7 | -139.5 |
| α | 7.16 | 7.16 | 8.73 | 8.99 | 10.72 | 5.84 |
| γ | 3.18 | 2.74 | 3.10 | 2.39 | 2.80 | 2.01 |
| | | | $M_3$ | | | |
| A | -210.7 | -170.6 | -139.1 | 9.6 | -1.0 | -85.0 |
| α | 5.90 | 6.06 | 7.29 | 7.59 | 9.13 | 4.80 |
| γ | 2.42 | 2.23 | 2.38 | 1.85 | 2.16 | 1.55 |
| | | | $M_4$ | | | |
| A | -177.7 | -132.5 | -91.9 | 79.3 | 74.3 | -41.8 |
| α | 4.40 | 4.57 | 5.56 | 5.90 | 7.12 | 3.67 |
| γ | 1.82 | 1.69 | 1.81 | 1.41 | 1.65 | 1.18 |

## 2.2. A share of surface defects in spatiotemporal phenomena

The catalytic $NO+H_2$ reaction on Rhodium exhibits kinetic and spatiotemporal oscillations observed by Field Emission Microscopy under particular conditions, where reaction (4) takes the key role [38, 60, 61]. The experiments were performed in the digital microscope described elsewhere [62 63]. Real-time images were recorded by a CCD camera. A 99.99% pure Rh wire 0.1 mm in diameter was used for the tip preparation by electrolytic etching. Spatiotemporal phenomena have been observed for the first time on the (100)- and (111)-oriented Rh tips. Later on, so as to study another area, the Rh wire under electrolytic etching was held not upright, but at the angle about 35° between (111) and (110) planes in a single crystal; a twin tip was resulted from this activity [31]. Visual observations have found out that the surface wave nucleation proceeds exactly at the grain boundary (GB) of that Rh twin tip in Fig. 7a, and the question comes why. The grain boundary is similar to a groove bottom on the reconstructed (110) plane, where atoms are depleted with the lost bonds in comparison with the smooth surface as represented schematically in Fig. 7b.

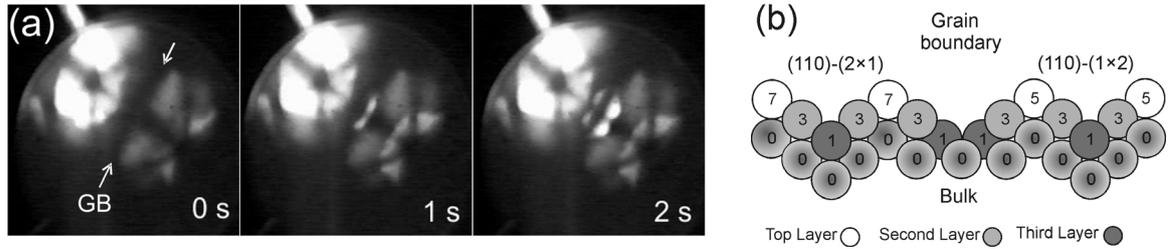

**Fig. 7 (a)** FEM screenshots of the 1 sec time lag exhibiting bright wave nucleation at the grain boundary of a Rh twin tip in the mixture of $P_{N0} = 1.1 \times 10^{-5}$ and $P_{H2} = 1.3 \times 10^{-4}$ Pa at 464 K (adopted from [31]); **(b)** The numbers of lost bonds around GB in comparison with the side views of fcc (110) plane.

The results of MIB calculations in Fig. 8 relate to Rh and show the peculiar GB location as compared to the perfect planes.

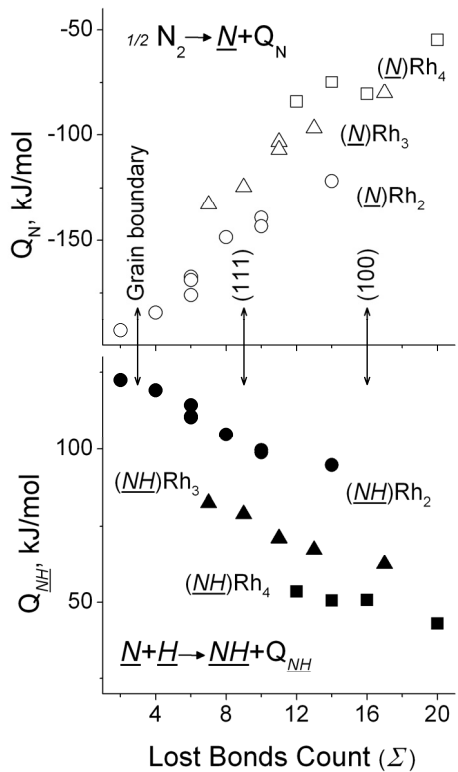

**Fig. 8** The heat of $N_2$ adsorption (empty symbols) and reaction (4) (filled symbols) vs. $\Sigma$ of Rh atoms involved in the double (circles), triple (triangles), fourfold (squares) bound $\underline{N}$ and $\underline{NH}$ formation, respectively, at basal Rh planes. The arrows indicate perfect (111), (100) terraces and the GB positions.

Following Eyring and Polanyi, the transition state under the relocation of the adsorbed species between equivalent sites includes the *half-broken* initial bond and the *half-formed* next one [64]. Then the activation energies of $\underline{N}$ and $\underline{NH}$ diffusion $E_{diff}$ can be estimated as the half-difference between the heats of adjacent states formation; it is true in case of the elementary move keeping the same number of bonds to surface (2 → 3; 6 → 7; 7 → 8); otherwise, $E_{diff}$ should be evaluated as the whole difference between relevant heats (1 → 2; 4 → 7; 5 → 8). Table 5 demonstrates that the adsorption sites with low $\Sigma$ around the grain boundary facilitate the $\underline{N}$ and $\underline{NH}$ species migration.

**Table 5** The heats of reactions (2), (3), (4) and activation energy of $\underline{N}$ and $\underline{NH}$ diffusion at the groove bottom $M_i$ sites of the Rh(110)-(1×2) plane (kJ/mol)

| No. | Site | $\Sigma$ | $Q_N$ | $Q_{\underline{NH}}$ | $Q_{NH}$ | Move Nos. | $E_{diff}(\underline{N})$ | $E_{diff}(\underline{NH})$ |
|---|---|---|---|---|---|---|---|---|
| 1 | $M_3$ | 11 | -111.8 | 70.6 | -2.3 | $1 \rightarrow 2$ | 47.2 | 13.3 |
| 2 | $M_2$ | 8 | -159.0 | 104.6 | -15.6 | | | |
| 3 | $M_2$ | 6 | -170.8 | 110.0 | -22.0 | $2 \rightarrow 3$ | 3.9 | 2.1 |
| 4 | $M_2$ | 4 | -178.6 | 113.5 | -26.2 | $4 \rightarrow 7$ | 82.1 | 19.9 |
| 5 | $M_2$ | 2 | -176.7 | 114.2 | -23.6 | $5 \rightarrow 8$ | 94.4 | 28.1 |
| 6 | $M_1$ | 5 | -244.6 | 167.9 | -37.8 | $6 \rightarrow 7$ | 8.1 | 4.2 |
| 7 | $M_1$ | 3 | -260.7 | 175.8 | -46.1 | | | |
| 8 | $M_1$ | 1 | -271.1 | 180.6 | -51.7 | $7 \rightarrow 8$ | 5.2 | 2.8 |

Table 5, Table 6 and Fig. 8 justify that the grain boundary enables the weakly bound $\underline{N}$ state, which is much more disposed to hydrogenation, and the equilibrium $\theta_{NH}$ by 5-8 orders of magnitude larger than in case of perfect terraces. These peculiarities make the grain boundary an active center for surface wave nucleation via reaction (4). As follows from Table 6, the high mobility of $\underline{N}$ and $\underline{NH}$ species facilitates the reaction (4) and the wave propagation in agreement with the total pattern.

**Table 6** Thermodynamic properties of Rh grain boundary in comparison with the terraces: the heats of indicated reactions; relative equilibrium coverage $\theta_{NH}$; diffusion activation energy $E_{diff}$ and relative diffusivity $D_{diff}$ for $\underline{N}$ and $\underline{NH}$ species (kJ/mol)

| | Rh$_4$(100) | Rh$_3$(111) | Rh$_4$(110) | Rh$_3$(110) | Rh$_2$(110)-(1×2) /GB/ |
|---|---|---|---|---|---|
| ½ N$_2$ → $\underline{N}$ + Q$_N$ | -56.9 | -110.3 | -84.1 | -106.9 | -178.6 |
| ½ N$_2$ + ½ H$_2$→$\underline{NH}$ + Q$_{NH}$ | 22.5 | -0.8 | 6.3 | 1.2 | -26.2 |
| $\underline{N}$ + $\underline{H}$ → $\underline{NH}$ + Q$_{\underline{NH}}$ | 40.5 | 70.7 | 51.5 | 69.2 | 113.5 |
| *$\theta_{NH}$ | $7.7 \cdot 10^{-9}$ | $1.8 \cdot 10^{-5}$ | $1.3 \cdot 10^{-7}$ | $1.2 \cdot 10^{-5}$ | 1 |
| $E_{diff}(\underline{N})$ = ½$\Delta$Q$_N$ | 100.3 | 58.3 | 59.1 | 36.3 | 3.9 |
| *$D_{diff}(\underline{N})$ | $1.9 \cdot 10^{-11}$ | $9.0 \cdot 10^{-7}$ | $7.3 \cdot 10^{-7}$ | $2.5 \cdot 10^{-4}$ | 1 |
| $E_{diff}(\underline{NH})$ = ½$\Delta$Q$_{NH}$ | 37.3 | 20.5 | 12.7 | 7.8 | 2.1 |
| *$D_{diff}(\underline{NH})$ | $1.2 \cdot 10^{-4}$ | $9.0 \cdot 10^{-3}$ | $6.6 \cdot 10^{-2}$ | 0.2 | 1 |

*According to the Boltzmann' estimation at characteristic oscillation temperature 470 K [31] and the same pre-exponential factors

Similar to Table 5, evaluations [38] have been carried out for each substrate in Fig. 5. For example, the difference 104.4 kJ/mol between $Q_N(M_3)$ and $Q_N(M_2)$ is an activation energy for *N* diffusion over Ru(0001) surface; then the energy for $N_2$ desorption 208.8 kJ/mol lies within the experimental range between 218 [65] and 190 kJ/mol [66].

### 2.3. Resonant active sites in catalytic ammonia synthesis

The $\Sigma$ model has been applied to catalytic $NH_3$ synthesis as follows. It is well known that $N_2$ dissociative adsorption is a rate-limiting step in the Haber–Bosch process. However, the product yield is determined by the combination of reactions (3) and (4) at equilibrium $\theta_N$. Then the opposite behavior of $Q_N(\Sigma)$ and $Q_{NH}(\Sigma)$ functions should result in a volcano curve, as a product of relevant equilibrium and/or kinetic constants, which maximum corresponds to the optimal catalytic center in ammonia synthesis. The further discussion considers extreme conditions of $\theta_N \to 0$; 1 as well as general cases of dynamic and equilibrium $\theta_N$, $\theta_{NH}$. Besides, the typical experimental pressure, reactant ratio and temperature ~800 K of $NH_3$ synthesis [28] allow to set $\theta_H << \theta_N$.

Extreme conditions reveal no useful information about catalytic centers. The maximal activity at the equilibrium $\theta_{NH}$ and $\theta_N \to 0$ requires $\alpha = \gamma$, while the opposite case of $\theta_N \to 1$ results in $\gamma = 0$.

The conditions of equilibrium $\theta_N$ are most efficient since MIB yields macroscopic characteristics; a set of enthalpies for reactions (2-6) is the only matter of consideration. If $N_2$ adsorption proceeds in equilibrium, then the first step of *N* hydrogenation becomes rate-limiting according to the elementary step constants [67, 68]. Therefore, the *NH* coverage and the rate of *NH* production can directly quantify the catalytic activity at the equilibrium and dynamic $\theta_{NH}$, respectively.

The maximum catalytic activity at *equilibrium* $\theta_{NH} = \dfrac{K_{NH}K_HK_N}{1+K_N}$ (where $K_H$, $K_N(\Sigma)$, $K_{NH}(\Sigma)$ are equilibrium constants; pre-exponential factors $K_{i,0}$ include $N_2$ and $H_2$ fluxes to the surface) means $\dfrac{\partial(\theta_{NH})}{\partial \Sigma} = 0$, which gives:

$$\theta_{NH} = const \cdot \frac{\exp\{-\gamma\Sigma / RT\}}{1 + K_{N,0}^{-1}\exp\{-(A + \alpha\Sigma) / RT\}} \tag{8a}$$

$$K_N = \frac{\alpha}{\gamma} - 1 \,; \ \theta_N = 1 - \frac{\gamma}{\alpha} \tag{8b}$$

The maximum catalytic activity at *dynamic* $\theta_{NH}$ corresponds to the maximum rate of $\underline{NH}$ production $\frac{\partial W_{NH}}{\partial \Sigma} = 0$ (where $W_{NH} = \frac{\partial(\theta_{NH})}{\partial t} = k_{NH} \frac{K_N K_H}{(1+K_N)^2}$; $k_{NH}$ is a rate constant for the reaction (4)), which gives:

$$\frac{\partial(\theta_{NH})}{\partial t} = const \cdot \frac{\exp\{\Sigma(\alpha - \gamma/2)/RT\}}{(1+K_{N,0}\exp\{(A+\alpha\Sigma)/RT\})^2} \tag{9a}$$

$$K_N = \frac{\alpha - \beta_{NH}\gamma}{\alpha + \beta_{NH}\gamma}; \quad \theta_N = \frac{1}{2}(1 - \frac{\beta_{NH}\gamma}{\alpha}), \tag{9b}$$

where $\beta_{NH} = 0.8$ is a Brønsted-Evans-Polanyi coefficient for activation energy of the reaction (4) [55, 69], and so $E_{NH}(\Sigma) = E_{NH,0} - \beta_{NH} \cdot Q_{\underline{NH}}(\Sigma) = const + 0.8\gamma\Sigma$.

In case of non-equilibrium $\theta_N$, where step (3) is rate-limiting, the steady state condition $\frac{\partial(\theta_N)}{\partial t} = k_N(1-\theta_N) - k_{NH}K_H\theta_N(1-\theta_N) = 0$ results in $\theta_N = \frac{k_N}{K_H k_{NH}}$. The BEP correlation gives $E_N = const - \alpha\beta_N\Sigma$ and $E_{NH} = const + \gamma\beta_{NH}\Sigma$, then the maximal activity $\frac{\partial \theta_N}{\partial \Sigma} = 0$ corresponds to the ineffective relation $\alpha\beta_N = \gamma\beta_{NH}$ similar to the extreme case of $\theta_N \to 0$. This result is expected since the rate-limitation by $N_2$ dissociative adsorption implies the lack of $\underline{N}$ [54] that is hardly the case because of the high activity of nitrides in $NH_3$ synthesis [70-72] and $N_2$ pressures up to 100 MPa. Therefore, equilibrium $N_2$ adsorption seems to be an acceptable occurrence.

Substitutions $\alpha$ and $\gamma$ from Table 4 in (8b), (9b) result in $K_N$, and then according to conventional $K_N = K_{N,0}\exp\{Q_N(\Sigma)/RT\}$ the optimal $\Sigma$ can be found as:

$$\Sigma = (-A + RT\ln K_N/K_{N,0})/\alpha$$

Pre-exponential factor $K_{N,0} = 0.027$ was estimated for typical experimental conditions of ammonia synthesis including the $N_2$ flux temperature ~800 K; $P_{N2}$ ~7 MPa; the catalyst temperature T ~800 K, atomic density $N_0 \sim 1\times10^{15}$ at cm$^{-2}$ and sticking coefficient for $N_2$ adsorption ~1. The values of $N_0 = 1.4\times10^{15}$ at cm$^{-2}$ and $K_{N,0} = 0.023$ are taken for Fe bcc planes. The typical error of 5% in $A$, $\alpha$ and $\gamma$ determination from linear fittings in Fig. 5 did not affect the final results. The relevant results of calculations are presented in Table 7, where steric limits follow from geometrical reasons in Fig. 3, and the equilibrium $\theta_{NH}$ at fcc Fe$_4$ site is a normalization unit of relative activity [32].

**Table 7** Optimal sum of the lost bonds $\Sigma$ at equilibrium and dynamic (eq/dyn) <u>$NH$</u> coverage and relative activity in $NH_3$ synthesis for $n$-atomic catalytic centers $M_n$ at basal fcc planes.

| | Pt | Rh | Ir | Ru | Re | Fe fcc |
|---|---|---|---|---|---|---|
| | $M_2$ (steric limits $2 \le \Sigma \le 14$) | | | | | |
| *$\Sigma$; eq/dyn | 37.8/36.9 | 33.4/32.5 | 25.0/24.2 | <u>11.6/10.6</u> | <u>10.9/10.0</u> | 28.8/27.5 |
| Activity | 5.7 | 41.5 | 158.9 | 1101.0 | 2269.8 | 60.1 |
| | $M_3$ (steric limits $5 \le \Sigma \le 13$) | | | | | |
| $\Sigma$; eq/dyn | 40.3/39.3 | 32.8/31.7 | 23.1/22.1 | 2.8/1.6 | 3.5/2.5 | 23.8/22.2 |
| Activity | 2.1 | 5.1 | 16.0 | 22.5 | 39.5 | 5.3 |
| | $M_4$ (steric limits $8 \le \Sigma \le 20$) | | | | | |
| *$\Sigma$; eq/dyn | 46.5/45.2 | 35.1/33.7 | <u>21.7/20.4</u> | -8.2/-9.9 | -6.1/-7.5 | <u>19.3/17.2</u> |
| Activity | 0.5 | 1.1 | 2.7 | 1.9 | 2.8 | 1 |

*$\Sigma$ Values meeting the accessible structures are underlined

Normalized catalytic activities are plotted according to Eq. (8a) and (9a) in Fig. 9, where volcano curves reflect general concept about the optimal <u>$N$</u> bond strength as the key factor for the best catalytic promotion [57]. Right sides of the pictures in Fig. 9 demonstrate more narrow dependencies, but relative curves are close to each other, and so the optimal $\Sigma$ slightly depends on the equilibrium or dynamic $\theta_{NH}$ behaviour (Table 7). Each sharp maximum in Fig. 9 follows from a given combination of structure and empirical parameters of a site similar to resonant circuit exhibiting natural frequency. The enormous advantage of the specific center over others in catalytic activity emphasizes the resonant $\Sigma$ character.

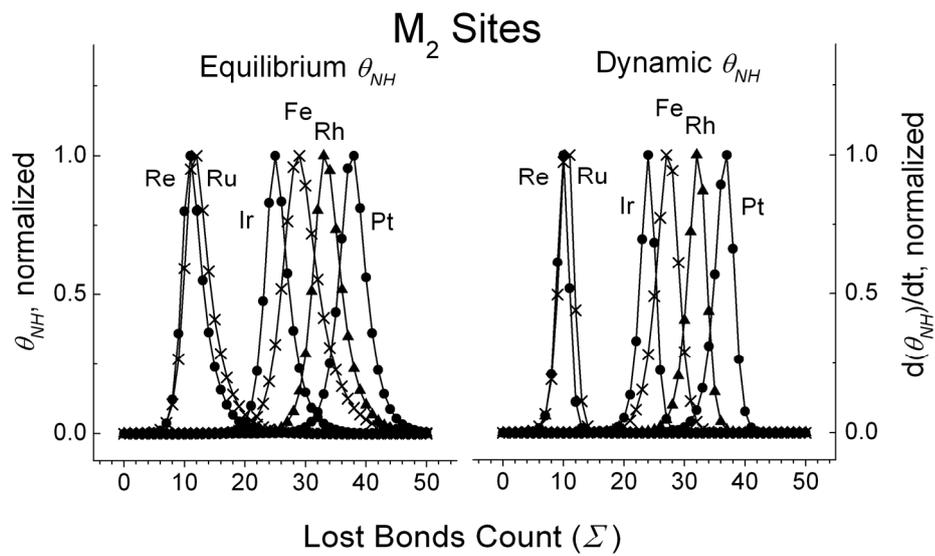

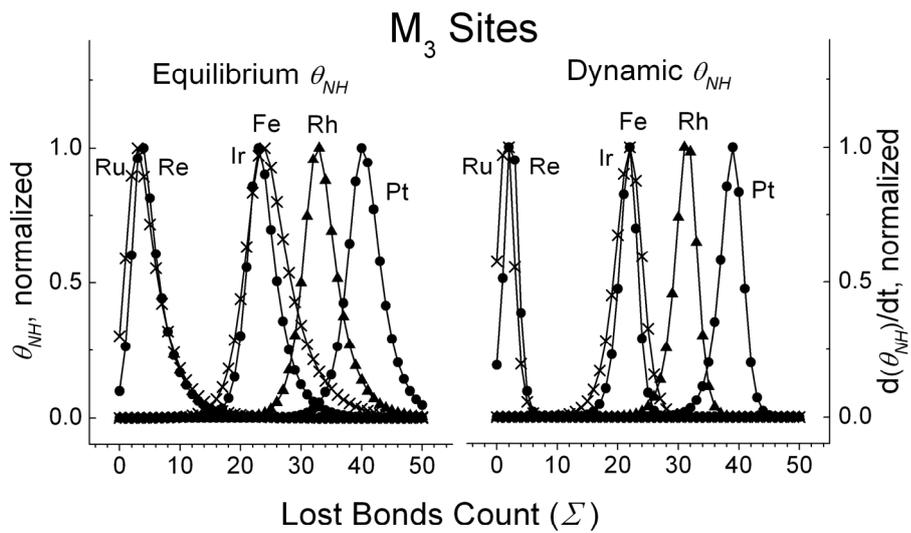

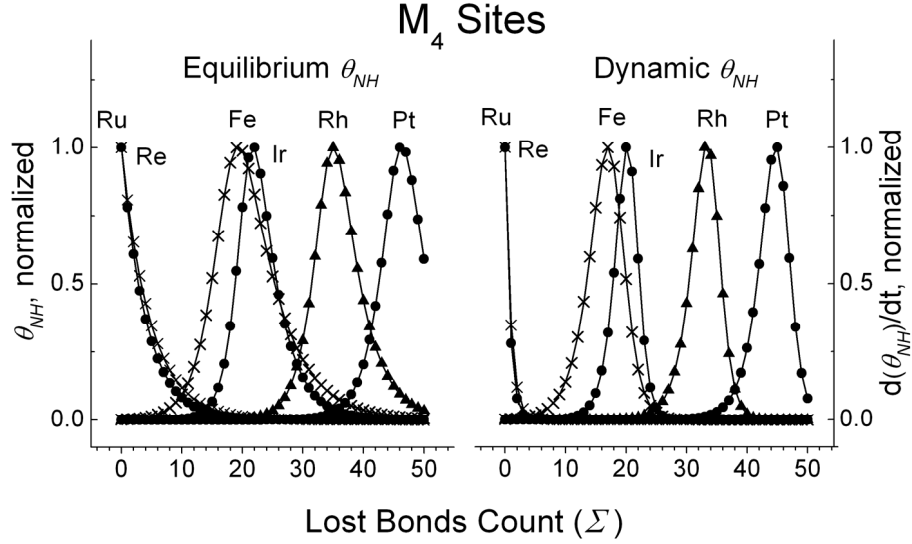

**Fig. 9** Normalized activity of 2-, 3- and 4-atomic centers on fcc planes of Pt, Ir, Re (circles); Fe, Ru (crosses) and Rh (triangles) in catalytic ammonia synthesis at equilibrium and dynamic _NH_ coverage.

The integers, nearest to fractional $\Sigma$ in Table 7, answer steric limits for $Ru_2$, $Re_2$ and $Ir_4$ only, whereas other centers require unreal $\Sigma$ or crystal structure (in case fcc $Fe_4$). Abnormally large $\Sigma$ for Pt, Rh and Ir (except $Ir_4$) mean that optimal sites for the ammonia synthesis are not available at perfect planes. Meanwhile, aforementioned $M_{2-4}$ clusters enable $\Sigma$ values up to 40. Therefore, the $\Sigma$ model predicts an extraordinary catalytic activity of small Pt, Ir and Rh ensembles, free or supported.

As follows from Fig. 3, the resonant $\Sigma$ = 10-12 for $Ru_2$ or $Re_2$ center can be found on fcc (110) plane (1×1, 1×2 or 2×1) similar to hcp ($1\bar{1}20$) plane of the real structure. In similar way, both $Ir_4$ centers ($\Sigma$ = 20-22) and hypothetical fcc $Fe_4$ centers ($\Sigma$ = 17-20) are available on fcc (100), (110)-(1×1) and (110)-(2×1) planes. Equilibrium $\theta_N$ values determined from Eqs. (8b), (9b) are within 0.4-0.6 and 0.2-0.4 ML at equilibrium and dynamic $\theta_{NH}$, respectively, for each site in Table 7. Resonant $\Sigma$ for dynamic $\theta_{NH}$ is usually smaller than for equilibrium $\theta_{NH}$ in Table 7 by ~1, thus giving preference to the reaction (4).

Table 7, Fig. 5 and Fig. 9 show that the Fe fcc sites are in a medium position between Ir and Rh. In contrast to fcc, Fe bcc planes exhibit the same $Q_N(\Sigma)$ and $Q_{NH}(\Sigma)$ dependences (Fig. 10a) at *any* number of atoms in a site. It is not entirely clear whether it comes from the bcc

structure itself or from the specific combination of parameters for that structure in Table 2. Similar analysis has revealed resonant $\Sigma = 12.4$ and $12.2$ for catalytic Fe center at equilibrium and dynamic $\theta_{NH}$, respectively. According to the experimental data [25], a "$C_7$ site" (where 7 is the coordination number, Fig. 10c) is responsible for the high catalytic activity of Fe(111) and (211) single crystals in comparison with others. Figures 10(b, c) demonstrate that $\Sigma = 13$ for "$C_7$ site" is inside the half-height width range of 11.3-13.5 predicted for resonant Fe site. Table 7 exhibits following relative activities of resonant centers $Fe_4(fcc)$ : $Ir_4$ : $Fe_{2-5}(bcc)$ : $Ru_2$ : $Re_2 = 1 : 2.7 : 3.4 : 1101 : 2270$, where $Fe_4(fcc)$ is a reference point and the subscript 2-5 for Fe(bcc) indicates the possible number of atoms in a site enabling integer $\Sigma$ nearest to the resonance. Dynamic $\theta_{NH}$ requires direct values of activation energies for the reaction (4) at each substrate and cannot be reliable for a similar comparison.

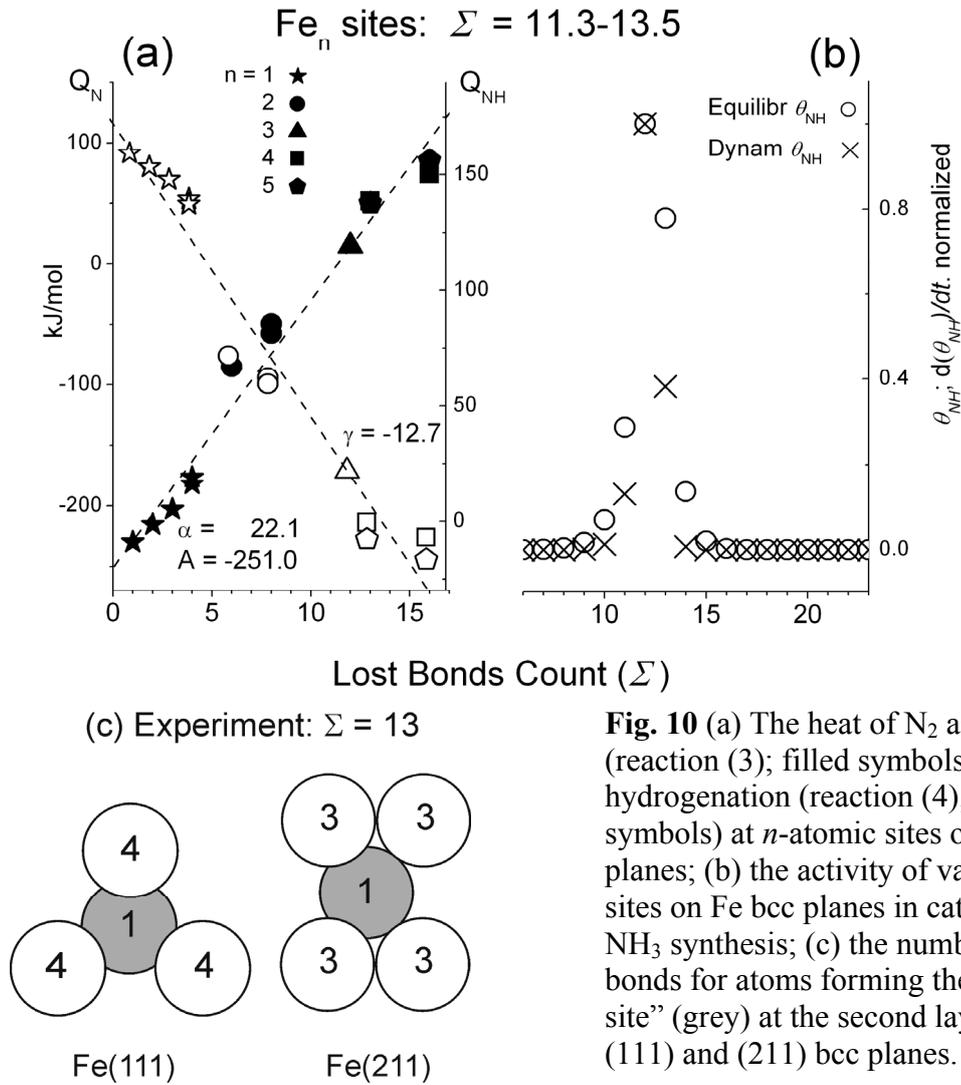

**Fig. 10** (a) The heat of $N_2$ adsorption (reaction (3); filled symbols) and $\underline{N}$ hydrogenation (reaction (4); open symbols) at $n$-atomic sites of Fe bcc planes; (b) the activity of various sites on Fe bcc planes in catalytic $NH_3$ synthesis; (c) the number of lost bonds for atoms forming the "$C_7$ site" (grey) at the second layer of (111) and (211) bcc planes.

There is a slight difference between activity ratios for fractional $\Sigma$ in Table 7 and integer $\Sigma$ in Table 8. Catalytic activity of the most active accessible sites corresponds to the following proportion $Pt_4 : Rh_4 : Fe_4(fcc) : Ir_4 : Fe_{2-5}(bcc) : Ru_2 : Re_2 = 4.8 \times 10^{-5} : 4.6 \times 10^{-3} : 1 : 2.7 : 3.4 : 1077 : 2269$. Resonant $\Sigma$ for Pt and Rh in Table 7 are banned by steric limits. Activity of the best accessible Pt and Rh sites increases in an unusual row of $M_3$, $M_2$, $M_4$ as $1 : 2.7 : 14.1$ and $1 : 1.9 : 28.8$, respectively.

**Table 8** Relative activities of $M_n$ sites in catalytic ammonia synthesis

| Site | $\Sigma$ | Activity | Site | $\Sigma$ | Activity |
|------|----------|----------|------|----------|----------|
| $Re_2$ | 11 | 2269 | $Ru_2$ | 12 | 1077 |
| $Re_2$ | 6 | 26.0 | $Ru_2$ | 6 | 15.8 |
| $Re_3$ | 9 | 8.9 | $Ru_3$ | 9 | 5.5 |
| bcc $Fe_{2-5}$ | 12 | 3.4 | bcc $Fe_{2-5}$ | 13 | 2.0 |
| $Ir_4$ | 22 | 2.7 | fcc $Fe_4$ | 19 | 1 |
| $Rh_2$ | 14 | $3.0 \times 10^{-4}$ | $Pt_2$ | 14 | $9.2 \times 10^{-6}$ |
| $Rh_3$ | 13 | $1.6 \times 10^{-4}$ | $Pt_3$ | 13 | $3.4 \times 10^{-6}$ |
| $Rh_4$ | 20 | $4.6 \times 10^{-3}$ | $Pt_4$ | 20 | $4.8 \times 10^{-5}$ |

Polycrystalline Re is by almost an order of magnitude more active in ammonia synthesis than the most active Fe(111) single crystal at reactant pressure 20 atm and temperature range of 603-713 K [73]. Table 8 demonstrates that $Re_2$ site is more active than $Re_3$ one. According to Fig. 3, a 3-atomic unit cell on Fe(111) holds a single site $Fe_{4,5}$ ($\Sigma$=12, 13), whereas similar cell holds 3/2 $Re_2$ sites on (0001) plane as a main component of polycrystalline surface. Then according to Table 8, $\Sigma$ model predicts 13.4 for Re-poly/Fe(111) activity ratio against "about an order" determined experimentally [73]. Similar evaluation gives 8.2 for Ru(0001)/Fe(111) activity ratio.

The rate of ammonia synthesis on single crystal Re catalysts at 20 atm reactant pressure and 720-900 K appeared to be very sensitive to surface structure. The revealed activity ratio is $1 : 94 : 920 : 2820$ for (0001), $(10\overline{1}0)$, $(11\overline{2}0)$, $(11\overline{2}1)$ crystal faces, respectively [21]. Such big difference is attributed to the greater level of openness and roughness of active surfaces. Relevant data in Table 9 demonstrate the activity ratio of $1 : 37 : 48 : 130$ for (0001), $(10\overline{1}0)$, $(11\overline{2}0)$, $(11\overline{2}1)$ planes, respectively.

**Table 9** Peculiarities of near-surface atoms and relative activity of catalytic centers on the relevant Re hcp planes at equilibrium $\theta_{NH}$

| Plane | Coordination number of surface atom | The number of lost bonds for surface atom | Integer $\Sigma$ for Re$_2$ site, by number of sites per unit cell | Relative activity of a single Re$_2$ site |
|---|---|---|---|---|
| $(11\bar{2}1)$ | 6; 8; 10; 11 in a $(C_6 \times C_6)$ unit cell | 6; 4; 2; 1 in a $(\Sigma_6 \times \Sigma_6)$ unit cell | $12 \times 1$ | 70 |
| | | | $10 \times 4$ | 70 |
| | | | $8 \times 3$ | 11 |
| | | | $7 \times 1$ | 3 |
| | | | $6 \times 1$ | 1 |
| | | | $5 \times 2$ | 0.3 |
| | | | $3 \times 1$ | 0.03; total 388 |
| $(11\bar{2}0)$ | 7; 11 in a $(C_7 \times C_7)$ unit cell | 5; 1 in a $(\Sigma_5 \times \Sigma_5)$ unit cell | $10 \times 2$ | 70 |
| | | | $6 \times 5$ | 1 |
| | | | $2 \times 2$ | 0.01; total 145 |
| $(10\bar{1}0)$ | 6; 10 in a $(C_6 \times C_6)$ unit cell | 6; 2 in a $(\Sigma_6 \times \Sigma_6)$ unit cell | $12 \times 1$ | 70 |
| | | | $8 \times 4$ | 11 |
| | | | $4 \times 1$ | 0.1; total 111 |
| $(0001)$ | 9 in a $(C_9 \times C_9)$ unit cell | 3 in a $(\Sigma_3 \times \Sigma_3)$ unit cell | $6 \times 3$ | 1; total 3 |

There is an admissible agreement between the experimental 3.1 and calculated 2.7 ratios for $(11\bar{2}0)$ and $(11\bar{2}1)$ activity, but experimental three orders difference in activity of $(0001)$ and $(11\bar{2}1)$ single crystals is much bigger in comparison with the data in Table 9. One order difference between the experimental and calculated $(10\bar{1}0)/(11\bar{2}0)$ activity ratio may result from the surface rearrangement in the course of reaction. For example, the *(2×1)*-like reconstruction of $(10\bar{1}0)$ plane will completely remove the most active Re$_2$ sites ($\Sigma = 12$) and thus drastically reduce activity. The evaluations similar to Table 9 have resulted in the activity ratio of 1 : 32 : 29 : 85 for Ru $(0001)$, $(10\bar{1}0)$, $(11\bar{2}0)$, $(11\bar{2}1)$ planes, respectively. The structure sensitivity of Re and Ru in NH$_3$ synthesis is in accordance with the experimental [26] and theoretical [27] data. One should note, however, that the present study considers clean metal surfaces only, whereas the nitride phase can also take active part in NH$_3$ production [70-72].

Ammonia synthesis over Ru-based catalysts is even more structure-sensitive than over Fe-based ones according to the studies of single crystals, supported catalysts, and DFT calculations [26, 65, 74, 75]. The superior activity of Ru-based catalysts is related to $B_5$ site, which consists of five Ru atoms exposing a three-fold hollow hcp site and a bridge site close to each other; these atoms are partly located on edges [26-28]. For Ruthenium, the integer closest to resonant $\Sigma = 11.6$ in Table 7 is available for $Ru_2$ centers only. Table 10 displays activities of $Ru_2$ centers provided by a single $B_5$ site (see figure inserted). The total activity of those sites and activity of $Ru_2$ sites on the (0001) plane differ by ~45 times, whereas a single $Ru_2$ site is ~70 times more active in comparison with that on the (0001) plane in Table 8.

**Table 10** Peculiarities of $Ru_2$ centers around a single $B_5$ site

| | Numbers of lost bonds for surface atoms | Integer $\Sigma$ for $Ru_2$ site, by number of sites per $B_5$ | Relative activity of a single $Ru_2$ site |
|---|---|---|---|
| $B_5$ site | 7; 3; 2 | $14 \times 2$ | 39.7 |
| | 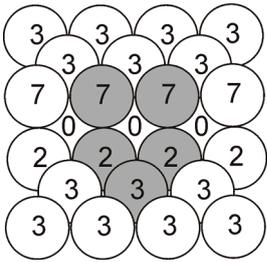 | $10 \times 4$ | 40.4 |
| | | $9 \times 2$ | 18.2 |
| | | $5 \times 2$ | 0.4; total 278.6 |
| (0001) | 3 | $6 \times 6$ | 1; total 6 |

There is another remark regarding the picture in Table 10. According to Fig. 5 and Table 4 the half heat of $N_2$ adsorption at the (100) step of $B_5$ site ($Ru_4$; $\Sigma = 18$) is 185.5 kJ/mol, while an exact $Q_N = 84.5$ kJ/mol for (0001) terrace ($Ru_3$; $\Sigma = 9$). These data enable to estimate the difference between activation energies for $N_2$ dissociation using a BEP coefficient of 0.8 as $0.8(2Q_N\{step\} - 2Q_N\{0001\}) = 1.67$ eV, close to 1.5 eV resulted from DFT calculations [75]. The valuable sensitivity of $\Sigma$ model to the local structure eliminates the BEP coefficients for individual steps [56].

Both acceptable agreements between calculated and relevant experimental data related to active sites in $NH_3$ synthesis and qualitative accordance with basic field concepts [25-28, 53] justify the $\Sigma$ model. It is identical to the approach implying a sum of coordination numbers in a site as a descriptor of the adsorption energy [76]. The similarity of the data in Fig. 5 and the behaviour of adsorption energies determined by DFT calculations [76] confirm reliability of both approaches. The $\Sigma$ model seems to provide more general description of imperfections and activities, because it can be successfully used for sites of different structure (fcc, bcc, hcp, clusters) and chemical composition (oxides, nitrides, alloys, etc.).

The present consideration concerns particular reactions, but it actually does not depend on substrate and adsorbate chemistry. Therefore, the structural sensitivity generally indicates that the heterogeneous catalytic reaction is rate-limited by interactions in the adsorbed layer including adsorption. The absence of structural sensitivity points at another rate-limiting stages like surface or bulk diffusion, reactant fluxes to surface, phase transitions, etc.

### 3. Oscillatory phenomena in NO+H$_2$ reaction

#### 3.1. The state-of-the-art

The oscillatory performance in the catalytic NO+H$_2$ reaction on noble metal surfaces is still a subject of discussion. To date, the reversible surface reconstruction, the coverage dependence of activation energies and the lack of vacant adsorption sites are considered as a respective feature responsible for oscillatory behaviour. However, each of these models has obvious limitations, though mathematical simulations of reaction kinetics demonstrate a good agreement with experimental data [77-83].

The reconstruction mechanism is based on strong difference between the activity of a pristine and reconstructed surface [78]. This model works well for the Pt(100) single crystal readily shifting the surface structure under adsorption, but it seems unusable for a good deal of Ir, Rh and Pt surfaces showing oscillations at invariable surface structure [30]. Moreover, the Pt(100) crystal also provides rate oscillations under certain conditions without visible reconstruction according to Low Energy Electron Diffraction observations [79].

Driving force of the vacancy model is a chain reaction in the adsorbed layer, where the rate of vacancy creation considerably exceeds the rate of consumption [19, 33]. It duly works for "surface explosion" at the saturated adsorbed layer and the lack of vacant states. However, the regular wave propagation is known to end with a nearly clean surface state, and there is no clear reason to restore the vacancy deficit. Therefore, this model can hardly explain the feedback mechanism of oscillations. The latter limitation has been overcome in the successful modelling of reaction rate oscillations provided by the coverage-dependent activation energies of key elementary steps, or conditioned by similar dependences of surface diffusion activation energies [80-83]. This is a clear trick without loss if simulations at fixed step constants do not reveal oscillations. Indeed, the tuning of correlation parameters enables to fit the calculated and experimental data, but this fitting is the only proof of initial model validation. Besides, the effect of strong lateral interactions on arrangements of the adsorbed particles is ignored in simulations [80], while kinetic equations directly depend on these probabilities [84]. Finally, the origin of wave nucleation and the chemical nature of traveling waves are still points for discussion in all aforementioned models [29, 85].

The importance of _NH_ species in oscillations during NO+H$_2$ reaction on noble metals was assumed previously [31, 36, 39, 45]. On the basis of experimental data and regularities disclosed by MIB calculations in a set of transition metal surfaces, the present study makes an attempt to clarify this viewpoint through comparative study of spatiotemporal phenomena on Rh tip [31] and modelling of the reaction kinetics on Pt(100) single crystal as the examples.

### 3.2. Reaction intermediates

The $NH_n$ species are inevitable intermediates for $NH_3$ production in the $NO+H_2$ reaction on noble metals [30]. Numerous studies have revealed stable $NH$ and $NH_2$ species at various surfaces of Pt [86], Rh [87], and Ir [88, 89]. For example, hydrazine decomposition includes a sequential formation of $NH_2$ and $NH$ on polycrystalline Ir foil [88], whereas the stable $NH$ and no sign of $NH_2$ species were found under $NH_3$ decomposition on Ir(110) [89]. Intermediate $NH$ particles were revealed in the titration reaction of $H_2 + NO$ on Pt(100) single crystal under identical experimental conditions by High Resolution Electron Energy Loss Spectroscopy [90] and Disappearance Potential Spectroscopy [91]. The "surface explosion" during $NO + H_2$ reaction on Rh tip was accompanied by a drastic work function ($\varphi$) drop $\sim 0.35$ eV compared to clean surface [92]. The similar $\varphi$ behaviour was observed for $NO+H_2$ interaction [77] and during sustained reaction rate oscillations on the Pt(100) single crystal [79]. The strong $\varphi$ decrease is most likely attributed to $NH_n$ species that therefore take a direct part in spatiotemporal performance. Thermal desorption from Rh filament, the support of Rh tip, has determined the stoichiometric ratio $n \sim 1$ for saturated $NH_n$ coverage [31].

Numerical studies testify an ability to form $NH_n$ species as the attribute of oscillations. Meanwhile, a good deal of the transition metal surfaces of Fe [93], Ru [94] and Re [95] are inactive in oscillations, but also form $NH_n$ species. The presence of $NH_n$ species seems to be a necessary, but insufficient condition of oscillatory behaviour. A comparative study has been performed in order to find out peculiarities of $NH_n$ particles at a *given plane,* which might be responsible for anomalous $NO+H_2$ reaction kinetics on a single crystal of the *same name*.

The fcc planes in Fig. 11 can be divided into two groups with respect to thermodynamic capability for reactions (5) and (6). It turned out that the $NH$ combination can proceed only on that surface, which exhibits oscillations in the $NO+H_2$ reaction, otherwise it is substantially endothermic. Reaction (6) is strongly endothermic for non-oscillating surfaces of Re and Ru, and strongly exothermic for oscillating surfaces of Pt, Ir, and Rh. Furthermore, Fig. 12 assigns the thermodynamic permissibility of reaction (5) on a given plane as a single feature consistent with the activity of similar single crystal in rate oscillations; the reactions (4), (6) are also enabled on this plane [36]. It is worth noting that Fig. 12 highlights pronounced activity of $NH$ species rather than importance of combination reaction.

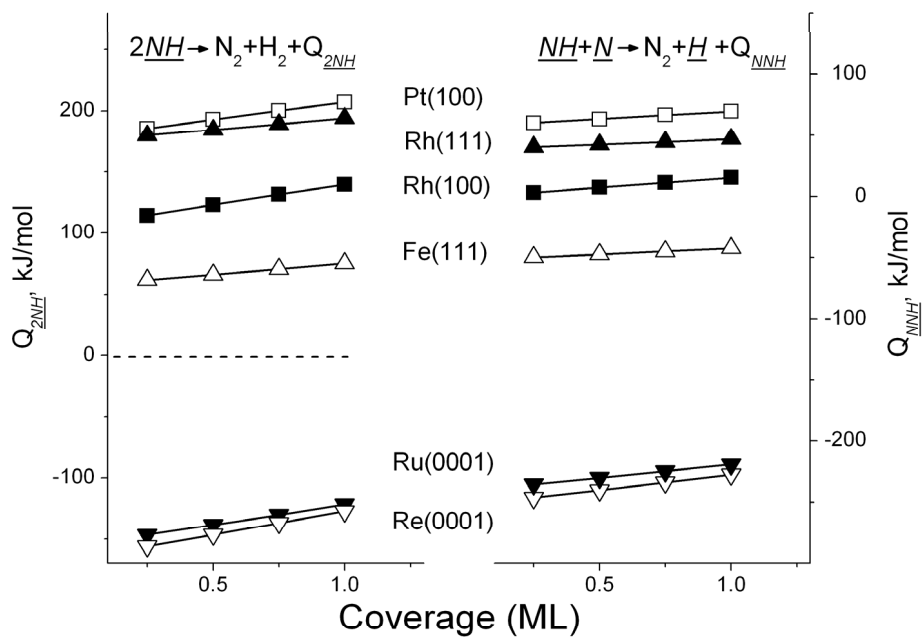

**Fig. 11** The heats of reactions (5) and (6) at indicated planes of Re (open triangles), Ru (filled triangles), Fe, Rh(100) (filled squares), Rh(111) (filled triangles) and Pt (open squares) vs. _NH_ coverage.

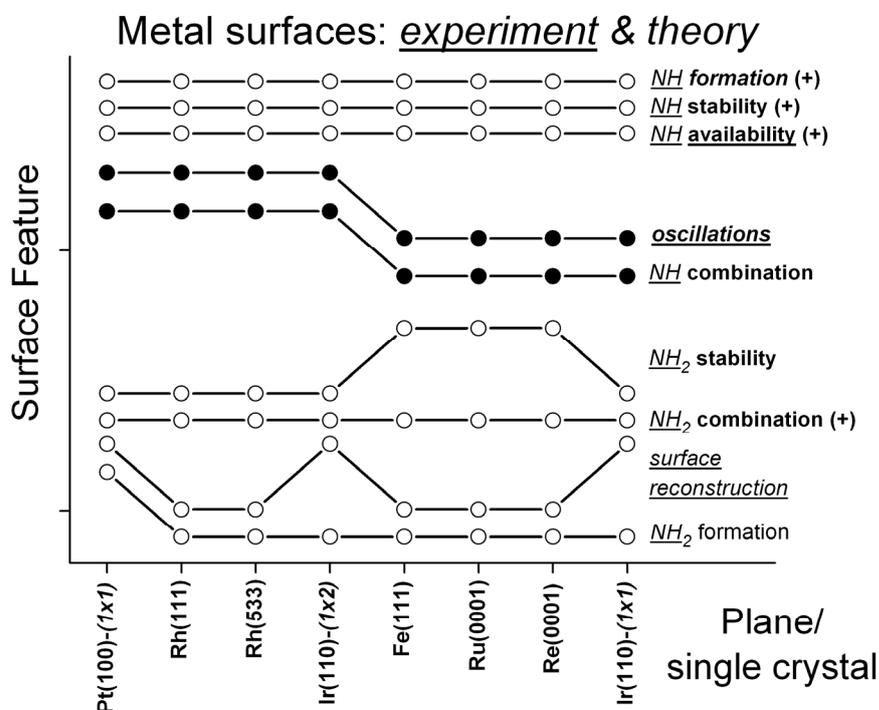

**Fig. 12** The correlation between experimental and calculated properties of metal surfaces. The higher points on broken lines or sign (+) at straight lines stand for the presence, and the lower points stand for the absence of a particular feature indicated at each curve.

The following regularities in Figs. 8, 11, 12 and Table 6 are found the most significant for understanding the mechanism of oscillation.

- There is a strong thermodynamic preference in reaction (3) at the grain boundary compared to perfect planes resulting in $10^5$-$10^9$ times difference in hypothetic equilibrium $\underline{NH}$ coverage at the middle temperature of regular oscillations 470 K [30]. This feature determines the spot on a surface for wave nucleation.

- Enormous $10^4$-$10^{11}$ range for $\underline{NH}$ and $\underline{N}$ diffusivity ratios implies the first may be regarded as *2D* gas around the fixed island of the second that is a key feature responsible for the island shape of travelling waves.

- The $\underline{NH}$ species at "oscillating" surfaces are well disposed towards the combination reaction (5) and $\underline{N}$ removal by step (6).

These items, in addition to necessary condition of oscillations, make possible to define a sufficient one: $\underline{NH}$ species should be highly mobile and reactive and follow "easy-come-easy-go" principle of ready formation and removal in the adsorbed layer. This regularity resembles the Sabatier principle [96] and uncovers the catalytic performance of $\underline{NH}$ species.

### 3.3. Spatiotemporal oscillations: qualitative consideration

The Scheme 1 gives a qualitative description of spatiotemporal phenomena on Rh twin tip in the NO+$H_2$ reaction [31].

**Scheme 1** The trigger model of oscillations in the NO+$H_2$ reaction on noble metals

*The permanent route*

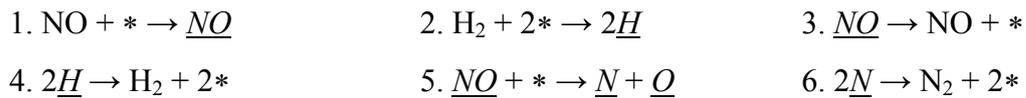

1. $NO + * \rightarrow \underline{NO}$    2. $H_2 + 2* \rightarrow 2\underline{H}$    3. $\underline{NO} \rightarrow NO + *$
4. $2\underline{H} \rightarrow H_2 + 2*$    5. $\underline{NO} + * \rightarrow \underline{N} + \underline{O}$    6. $2\underline{N} \rightarrow N_2 + 2*$

*The temporal route*

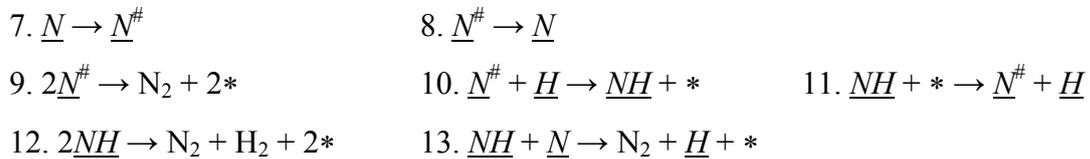

7. $\underline{N} \rightarrow \underline{N}^{\#}$    8. $\underline{N}^{\#} \rightarrow \underline{N}$
9. $2\underline{N}^{\#} \rightarrow N_2 + 2*$    10. $\underline{N}^{\#} + \underline{H} \rightarrow \underline{NH} + *$    11. $\underline{NH} + * \rightarrow \underline{N}^{\#} + \underline{H}$
12. $2\underline{NH} \rightarrow N_2 + H_2 + 2*$    13. $\underline{NH} + \underline{N} \rightarrow N_2 + \underline{H} + *$

*Common steps*

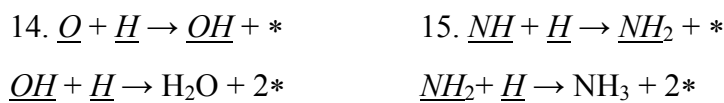

14. $\underline{O} + \underline{H} \rightarrow \underline{OH} + *$    15. $\underline{NH} + \underline{H} \rightarrow \underline{NH_2} + *$
$\underline{OH} + \underline{H} \rightarrow H_2O + 2*$    $\underline{NH_2} + \underline{H} \rightarrow NH_3 + 2*$

Here symbol * denotes the vacant site

The permanent route provides a steady state including adsorption and dissociation of the reactants, relevant interactions in the adsorbed layer and products desorption via common steps. The temporal pathway is responsible for oscillations; it consists of the steps ascertained from experimental observations (the way of nucleation and propagation of surface wave) and theoretical investigation (trigger steps 12, 13 in Sch. 1).

Characteristic conditions of oscillatory phenomena (T = 420-520 K; a large excess of $H_2$ in reaction mixture [30]) imply a strong $\underline{N}$ domination in the adsorbed layer and disable the bulk oxide formation. The competition between formation and desorption rates determines the current $\underline{N}$ coverage and the reaction rate. The kinetics of NO+$H_2$ reaction on Rh filament is characterized by two temperature regions separated by the breakpoint in Arrhenius plot [31]. The surface is saturated with $\underline{N}$ in lower-T region and reaction rate is limited by desorption, while the temperature rise removes $\underline{N}$ and reaction rate becomes limited by its formation. The Rh twin tip is cut out two adjacent monocrystalline grains separated by the borderline, and the same grains make up the filament [31]. Therefore, similar samples under identical conditions must undergo the same chemical interactions. Indeed, spatiotemporal oscillations on Rh tip spot-welded to filament occur exactly around the breakpoint in Arrhenius curve; the alteration of the total pressure or NO/$H_2$ ratio have resulted in the same temperature shift of that breakpoint and sustained traveling waves. The appropriate conditions enable the reserve reaction route for $\underline{N}$ removal (being the transition from lower-T to higher-T region), which occurs as the wave propagation. On the grounds of above mentioned findings, a cycle of spatiotemporal oscillations [31] can be interpreted within Sch. 1 as following.

- The wave propagation ends with a resting state of almost clean surface.

- Reaction proceeds on that surface by the permanent pathway without appreciable $\underline{NH}$ coverage; $\underline{N}$ atoms are randomly distributed (no islands within 20 Å FEM resolution were observed), and the rate of $\underline{N}$ formation exceeds the rate of desorption. As a result, the surface gradually fills up with $\underline{N}$ species up to critical coverage enabling the reserve reaction route, where $\underline{NH}$ species serve as catalysts for explosive $\underline{N}$ removal by steps 7-13.

- The temporal pathway starts with $\underline{N}^{\#}$ coverage growth at the grain boundary conditioned by the saturation of perfect terraces with the more strong bound $\underline{N}$ state.

- According to Fig. 8 and Table 6, the $\underline{N}^{\#}$ state has a sizeable thermodynamic preference in reaction (4) over the $\underline{N}$ state and encourages the growth of $\underline{NH}$ coverage, thus giving rise to surface wave nucleation via steps (10-13). In contrast to nitrogen atoms inside the $\underline{N}$ layer, $\underline{N}$ atoms at the border of layer have access to $\underline{H}$ species just released in step (13) or resulted from $H_2$ adsorption. The boundary atoms take the role of $\underline{N}^{\#}$ state and enable the next $\underline{NH}$ formation for a new $\underline{N}$ removal. The *2D $\underline{NH}$* gas "eats up" the fixed $\underline{N}$ island from the outside providing surface wave propagation, which closes a cycle of oscillations and leaves nearly clean surface in a starting state ready to set the next cycle. This pattern is similar to the $CO+O_2$ reaction on the Pt(100) single crystal, which was found to proceed along the boundary of $\underline{O}$ islands [97].

The critical $\underline{N}$ coverage necessary for initiation of the trigger pathway depends on external conditions ($T$, $P_{H2}$, $P_{NO}$). The quasi-equilibrium $\underline{N}$ saturation at the point of wave nucleation provides a small enough diffusion time for $\underline{NH}$ particle to undergo steps (12, 13), before it decomposes by step (11) in Sch. 1. The trigger origin and temporality of steps (7-13) explain the feedback mechanism of oscillations. The residual $\underline{NH}$ coverage behind the wave front is responsible for the $\varphi$ drop, i.e. the wave brightness and sharpness. The partial $\underline{NH}$ hydrogenation to form $NH_3$ is not very effective [30] and should not affect the oscillatory behaviour. The following points demonstrate a consistency between the double-route Scheme 1 and relevant experimental data on selectivity, influence of temperature, and work function change.

• Both the $H_2$ excess in reactant mixture and the reduced rate of $\underline{N}$ combination at low coverage make $NH_3$ production (15) more effective compared to step (6) in Sch. 1; on the contrary, an increase in $\underline{N}$ coverage will increase the rate of step (6) and decrease the rate of step (15). Therefore, $N_2$ production should be in counter-phase with $NH_3$ and $H_2O$ formation as it really takes place for the Rh and Ir single crystals [30]. On the other hand, the $\underline{NH}$ hydrogenation at Pt(100) plane is 10-30 kJ/mol more exothermic compared to Rh(111), Rh(335) and Ir(110) planes [39]. Then one can expect the phase alignment in rates of $N_2$ and $NH_3$ formation on Pt in accordance with experiments [30].

• The temperature lift increases the diffusivity of $\underline{NH}$ species and decreases the critical $\underline{N}$ coverage necessary for initiation of the temporal reaction pathway. It means the decrease in amplitude and increase in frequency of spatiotemporal oscillations on the temperature rise in agreement with experimental observations [98].

• The work function drop provided by a residual $\underline{NH}$ coverage behind the wave should be more pronounced under "surface explosion" against regular waves since $\underline{N}$ coverage is much larger in the former case. This consequence of the trigger model is also in line with experimental data [77, 79, 92].

### 3.4. Reaction rate oscillations: quantitative description

The Scheme 1 is further considered as a kinetic model. According to set of the adsorbed species {$\underline{H}$; $\underline{N}$; $\underline{N}^{\#}$; $\underline{NO}$; $\underline{NH}$; $\underline{O}$}, the temporal variations of partial coverages are defined by the ordinary differential equation (ODE) system $\Theta' = F(\Theta, t)$:

$$\frac{d\theta_H(t)}{dt} = 2R_2 - 2R_4 - R_{10} + R_{11} + R_{13} - 2R_{14} - 2R_{15} \qquad \frac{d\theta_{NO}(t)}{dt} = R_1 - R_3 - R_5$$

$$\frac{d\theta_N(t)}{dt} = R_5 - 2R_6 - R_7 + R_8 - R_{13} \qquad \frac{d\theta_{N^{\#}}(t)}{dt} = R_7 - R_8 - 2R_9 - R_{10} + R_{11}$$

$$\frac{d\theta_{NH}(t)}{dt} = R_{10} - R_{11} - 2R_{12} - R_{13} - R_{15} \qquad \frac{d\theta_O(t)}{dt} = R_5 - R_{14}$$

Here $R_i$ is the rate of corresponding elementary step in Scheme 1.

The elementary Langmuir-Hinshelwood (LH) reaction rates $R_1 - R_{15}$ for the above ODE system are expressed then:

$$R_1 = P_{NO}s_{NO}k_1\theta_{vac} \qquad R_2 = P_{H_2}s_{H_2}k_2\theta_{vac}^2 \qquad R_3 = k_3\theta_{NO}$$

$$R_4 = k_4\theta_H^2 \qquad R_5 = k_5\theta_{NO}\theta_{vac} \qquad R_6 = k_6\theta_N^2$$

$$R_7 = k_7\sqrt{\theta_N} \qquad R_8 = k_8\sqrt{\theta_{N^{\#}}} \qquad R_9 = k_9\theta_{N^{\#}}^2 \qquad (10)$$

$$R_{10} = k_{10}\theta_{N^{\#}}\theta_H \qquad R_{11} = k_{11}\theta_{NH}\theta_{vac} \qquad R_{12} = k_{12}\theta_{NH}^2$$

$$R_{13} = k_{13}\theta_N\theta_{NH} \qquad R_{14} = k_{14}\theta_O\theta_H \qquad R_{15} = k_{15}\theta_H\theta_{NH}$$

The vacancy coverage $\theta_{vac}$ is determined from monolayer normalization:

$$\theta_{vac} = 1 - \theta_{NO} - \theta_H - \theta_N - \theta_{N^{\#}} - \theta_O - \theta_{NH}$$

The rate of gaseous $N_2$ production (ML/s) was taken as a total reaction rate:

$$R_{N_2} = k_6\theta_N^2 + k_9\theta_{N^{\#}}^2 + k_{12}\theta_{NH}^2 + k_{13}\theta_N\theta_{NH}$$

The steps (14) and (15) in Sch. 1 are considered as rate limiting for the $H_2O$ and $NH_3$ production, respectively [80]. The square root dependence for $R_7$ (10) directly results from experimental observations of traveling waves in Fig. 1 and represents the $\underline{N}^{\#}$ state formation along the border of $\underline{N}$ island. The formation of minor reaction product $N_2O$ was omitted and $k_{1-15}$ values were fixed in the further simulations.

Numerical integration of the ODE system $\Theta' = F(\Theta, t)$ has been performed by the Euler method [99]. This assumes the current coverage set $\{\Theta_i(t_i)\}$ at the time point $t_i = t_0 + ih$, $i = 1$, 2, ..., $n$ and starting conditions $\{\Theta(t_0)\} = \{\Theta_0\}$ can be approximated as $\Theta_{i+1} = \Theta_i + hF(\Theta_i, t_i)$, where $h = 1.12 \cdot 10^{-3}$ s is an integration time step determined from the simulated kinetics of NO adsorption on clean surface $\theta_{NO} = \theta_{NO}(h)$ and compared to the monomolecular LH adsorption $\theta_{NO} = \theta_{NO}(t)$ [38]. Operating $h$ seems quite suitable for continuous description of the reaction kinetics at experimental pressure range $10^{-4}$-$10^{-3}$ Pa [30] accepted in the simulations.

Regular oscillations in partial coverages and $N_2$ production were obtained with fixed step constants in Table 11. Each of the $\Delta k_i$ ranges is compared with an Admissible Limit (AL) to verify the reality of operational step constants. The AL was calculated by Arrhenius equation $k_i = k_{i,0} \exp\{-E_i/RT\}$ using the reference data for $k_{i,0}$ and $E_i$ within the temperature range of 420-520 K characteristic for regular oscillations on Pt(100). The largest number of reference data for the Pt(100) single crystal among others related to noble metals is the only reason to use this surface in testing of the simulation parameters [30].

**Table 11** The range of step constants $\Delta k_i$ in Scheme 1 enabling regular oscillations in Fig. 13; reference $k_{i,0}$ and $E_i$ values; Admissible $k_i$ Limit (AL) for the temperature range 420-520 K relevant to the Pt(100) single crystal

| $k_i$ | $\{\Delta k_i\}$, (s·ML)$^{-1}$ | $k_{i,0}$, s$^{-1}$ | $E_i$, kJ/mol | AL $k_i$ (s·ML)$^{-1}$ | Comment |
|---|---|---|---|---|---|
| $k_1$ | 0.2347 | 2.12·10³ Pa$^{-1}$ 300 K, $s = 1$ | 0 | - | $s = 0.9$ [100] |
| $k_2$ | 6.64 | 8.20·10³ Pa$^{-1}$ 300 K, $s = 1$ | 0 | - | $s = 0.21$ [101] $s = 0.48$ [102] |
| $k_3$ | 0.02978-0.0299 | 1.7·10¹⁵ | 154.8 | 1·10$^{-4}$-0.5 | [80] |
| $k_4$ | 8.76-8.83 | 1·10¹² | 104.6 | 0.1-29.8 | [80] |
| $k_5$ | 1.915-1.921 | 2·10¹⁵ 2.1·10¹² | 117.2 37.7-192.5 116.7 | 4.9-3·10³ 1·10$^{-4}$ - 4·10¹⁰ 6.4·10$^{-3}$-3.9 | [80] [103] [79] |
| $k_6$ | (0-6)·10$^{-5}$ | 1·10¹³ | 100.4 142.3 140-170 | 3-800 1·10$^{-5}$ - 5·10$^{-2}$ 1·10$^{-8}$ - 0.1 | [80] [103] $2E_{diff}(\underline{N})$ [45] |
| $k_7$ | (9.79-9.83) 10$^{-3}$ | 10⁸-10¹¹ | 70-85 | 3·10$^{-3}$ - 1·10² | $D_0 (\underline{O})$ [104] $E_{diff}(\underline{N})$ [45] |
| $k_8$ | $k_8/k_7 = 0$-0.25 | | | | |
| $k_9$ | (1-5.7)·10$^{-5}$ | 1·10¹³ | 140-170 | 1·10$^{-8}$ - 0.1 | [80] $2E_{diff}(\underline{N})$ [45] |
| $k_{10}$ | 0.507-0.512 | 1·10⁹ | 62.8 70-85 | 3·10$^{-2}$ - 5·10² | [80] $E_{diff}(\underline{N})$ [45] |
| $k_{11}$ | 0-0.03 | 1·10¹³ | 121.3 121-143.2 | 1·10$^{-2}$ - 6.5 1·10$^{-5}$ - 7.0 | [80] $E_9 + Q_{\underline{NH}}$ [45] |
| $k_{12}$ | 0.98-1.12 | 1·10⁹ 1·10¹³ | 74.1 100.4 | 0.6-800 | [80,83] |
| $k_{13}$ | 4.70-5.09 | 1.4·10¹³ | 90-105 | 1.2-1·10⁴ | $\nu_{PtN}$ [90] $E_{diff}(\underline{NH})$ + $E_{diff}(\underline{N})$ [45] |
| $k_{14}$ | 1.082-1.10 | 1·10¹³ 1.7·10¹⁰ 5.2·10⁷ | 54.4 71.1 | 8·10$^{-2}$-2·10⁷ | [80] [103] [105, 106] |
| $k_{15}$ | 0-0.10 | 1·10⁹ 2·10⁸ | 74.1-79.5 | 3·10$^{-2}$-36.0 | [80] $D_0 (\underline{H})$ [83,107] |

The $k_1$, $k_2$ constants stand for NO and $H_2$ fluxes to the surface and do not depend on temperature. The accepted $k_1$ and $k_2$ values at sticking coefficients $s_{NO}$, $s_{H2} = 1$ correspond to $P_{NO} = 1.1 \cdot 10^{-4}$ and $P_{H2} = 8.1 \cdot 10^{-4}$ Pa, respectively. The decrease of $s_{NO}$, $s_{H2}$ within the range of reference data [100-102] means an increase of virtual partial pressure and/or the variation of $H_2/NO$ ratio within 7 – 30, inside the experimental ratio 0.5 – 30 [30].

The operating $\Delta k_{3-5}$ and $\Delta k_{14}$ ranges comply with the AL data [80, 103, 105, 106].

The parameter $E_6$ can not be determined correctly from the experimental data since $\underline{N}$ desorption from the Pt(100)-$(1 \times 1)$ surface is accompanied by $(1 \times 1)$-to-$(hex)$ reconstruction.

The double diffusion activation energy of nitrogen atom $E_{diff}$ ($\underline{N}$) was taken as $E_6$, and $E_{diff}$ ($\underline{N}$) was calculated by MIB [45].

No reference data have been found for $k_7$ parameters. The present evaluation supposes that the $\underline{N}^{\#}$ state formation is limited by $\underline{N}$ diffusion; and so $E_{diff}$ ($\underline{N}$) was accepted for $E_7$ and pre-exponential factor for $\underline{O}$ diffusion was taken as $k_{7,0}$.

The activation energy of the $\underline{NH} + \underline{H} \rightarrow \underline{NH_2}$ reaction was considered as a lower limit of $E_{12}$, whereas the higher limit should not be above the activation energy for $\underline{N}$ combination by step (6) in Sch. 1 [80].

The parameters $E_9$ and $E_{13}$ were respectively estimated by the sum of diffusion activation energies of reaction components as $2E_{diff}$ ($\underline{N}$) and $\{E_{diff}$ ($\underline{NH}$) $+ E_{diff}$ ($\underline{N}$)$\}$ [45]. Pre-exponential factor $k_{13,0}$ corresponds to the $\nu_{PtN}$ valence mode determined experimentally for $\underline{NH}$ species at the Pt(100)-$(1 \times 1)$ single crystal surface [90].

The parameters $E_{10}$ and $E_{11}$ are not confirmed in ref. [80] anyhow; the interval of $E_{diff}$ ($\underline{N}$) [45] was accepted for $E_{10}$ since the step (10) in Sch. 1 assumes $\underline{N}$ displacement from the strongly bound state, then the sum $E_{10}$ and the enthalpy of step (10) gives $E_{11}$ parameter.

The cited $k_{15,0}$ and $E_{15}$ values were reported among others [80] as empirical parameters of the best fit between the calculated and experimental data. Pre-exponential factor $D_0$ for $\underline{H}$ diffusion [107] is another $k_{15,0}$ estimation taking account for exceptional $\underline{H}$ mobility.

Above consideration validates that each of step constant ranges $\{\Delta k_i\}$ in Table 11 enabling regular oscillations is inside the Admissible Limit and thereby quite real and acceptable.

The NO+H$_2$ reaction reveals oscillatory behavior under certain experimental conditions providing optimal partial coverages [30]. The overage set $\{\Theta_0\}_1$ in Table 12 is typical for the time point of wave nucleation, when the surface is mainly covered with $\underline{N}$ and small amount of $\underline{NO}$ and $\underline{O}$ species [35].

**Table 12** Starting coverage sets $\{\Theta_0\}_1$ and $\{\Theta_0\}_2$ used in simulations of Fig. 13, Fig. 14(a, b, d) and Fig. 14c, respectively

| Surface species | *Vacancy* | $\underline{H}$ | $\underline{N}$ | $\underline{N}^{\#}$ | $\underline{NO}$ | $\underline{NH}$ | $\underline{O}$ |
|---|---|---|---|---|---|---|---|
| $\{\Theta_0\}_1$, ML | 0.05 | 0.05 | 0.75 | 0 | 0.05 | 0 | 0.1 |
| $\{\Theta_0\}_2$, ML | 0.4 | 0 | 0.6 | 0 | 0 | 0 | 0 |

Figure 13 shows the results of numerical modelling of reaction kinetics within the Sch. 1 using average step constants $\{\Delta k_i\}$ in Table 11 (except $k_8 = k_{15} = 0$), the starting coverage set $\{\Theta_0\}_1$ in Table 12, and an invariant integration step $h = 1.12 \cdot 10^{-3}$ s.

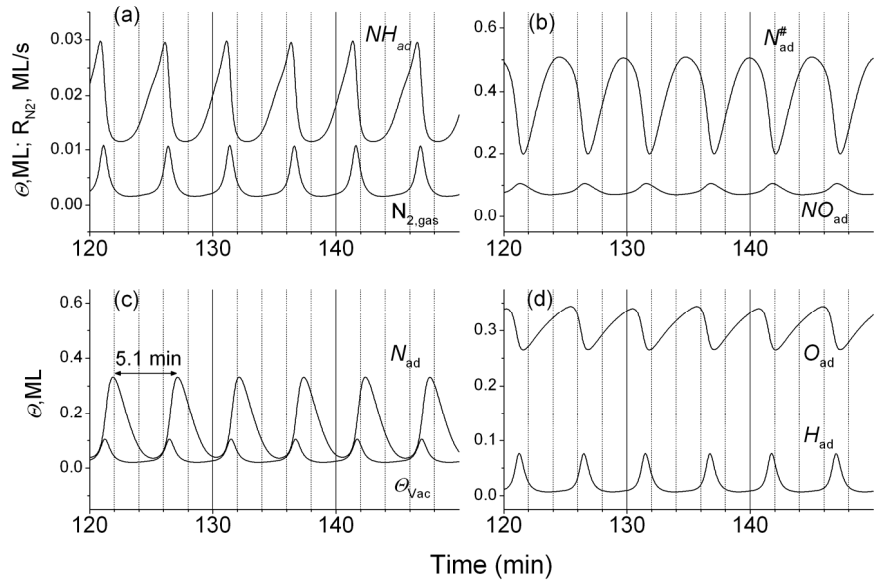

**Fig. 13** Modelled regular oscillations in coverage of (a) $\underline{NH}$, and $N_2$ production rate; (b) $\underline{N}^\#$ and $\underline{NO}$; (c) $\underline{N}$ and vacancies $\theta_{vac}$; (d) $\underline{O}$ and $\underline{H}$ during NO+$H_2$ reaction on Pt(100) at $P_{NO} = 1.1 \cdot 10^{-4}$, $P_{H2} = 8.1 \cdot 10^{-4}$ Pa; average step constants $\{\Delta k_i\}$ in Table 11, except $k_8 = k_{15} = 0$; starting coverage set $\{\Theta_0\}_1$ in Table 12; invariant integration step $h = 1.12 \cdot 10^{-3}$ s.

The addition of step (15) to simulations in Sch. 1 results in $NH_3$ (phase coupled with $N_2$) formation and shows the same oscillatory behaviour. It is worth noting that the total rate of $N_2$ production ~ 0.01 ML/s in Fig. 13(a) is inside the experimental range of 0.008-0.030 ML/s determined on Pt(100) single crystal at the same partial pressure $P_{NO} = 1.1 \cdot 10^{-4}$ Pa [79, 100].

In spite of thermodynamic preference, the reverse step (8) in Sch. 1 was omitted under modelling since the forward step (7) is conditioned by the $\underline{N}$ saturation, and so there are no vacancies for $\underline{N}$ allocation. Besides, the surface species may migrate from strongly to weakly bound states to form the reaction-diffusion wave as it has been observed for $\underline{CO}$ oxidation at the O-saturated Ag/Pt(110) surface [108]. However, the testing simulations have revealed the sustained rate oscillations over the range 0-0.25 of $k_8/k_7$ ratio. Table 11 displays the sensitivity of oscillations to step constants. The most $k_i$ can be changed within 0.2-15 %, while $k_6$, $k_9$ and $k_{11}$ can be varied over much wider range maintaining the regularity of oscillations.

The influence of operational parameters on oscillatory features in Fig. 13 is shown in Fig. 14. As an example, Fig. 14(a) shows fast decay of oscillations ending with the steady state as a result of the $k_{13}$ replacement from 5 to 4 under regular oscillations. The opposite $k_{13}$ change to 5.15 reveals in Fig. 14(b) the amplitude beating and frequency disturbance followed by the abrupt decrease in total reaction rate and domination of $\underline{O}$ and $\underline{N}^{\#}$ coverage. Oxygen displaces eventually the $\underline{N}^{\#}$ state, then surface becomes completely poisoned with $\underline{O}$ and the reaction stops. The variation of another step constant beyond the $\{\Delta k_i\}$ ranges in Table 11 influences the reaction kinetics in a similar way.

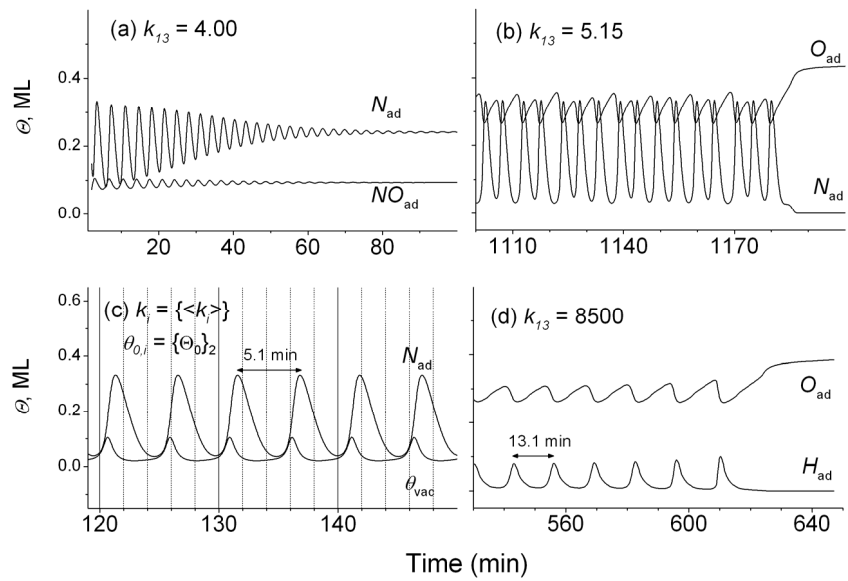

**Fig. 14** Regular oscillations in Fig. 13 affected by the variation of operational parameters: replacement of $k_{13} = 5.00$ by (a) $k_{13} = 4.00$ (changeover to the steady state) and (b) $k_{13} = 5.15$ (complete reaction inhibition); (c) replacement of starting coverage set $\{\Theta_0\}_1$ by $\{\Theta_0\}_2$ (phase shifted oscillations); (d) unstable oscillations at a distant $k_{13} = 8500$.

The comparison of Fig. 13(c) and Fig. 14(c) clarifies the influence of starting conditions on oscillations. The replacement of $\{\Theta_0\}_1$ by $\{\Theta_0\}_2$ coverage set in Table 12, retaining the step constants, does not affect oscillations, but reveals the phase shift called forth by different time required for reaching the similar partial coverages. This is a direct evidence for the stability and attractive quality of the oscillatory reaction mode within a set of $\{\Delta k_i\}$ ranges in Table 11. The $\{\Delta k_i\}$ set may be not the only one of this kind. Figure 14(d) shows unstable oscillations at a distant $k_{13} = 8500$ compared to operational range $\Delta k_{13} = 4.70$-$5.09$ (s·ML)$^{-1}$ in

Table 11. These oscillations stopped suddenly after about 50 regular cycles and the surface became poisoned with $\underline{O}$ as in the case of Fig. 14(b).

The conventional mechanism of the NO+$H_2$ reaction does not reveal oscillations and the previous models have had to use the conceptual dependences of step constants on coverage to fit experimental data [80-83, 109, 110]. The introduction of weakly bound, but very active $\underline{N}^{\#}$ state at the surface defects provides the quantitative description of rate oscillations at *fixed* step constants. This approach conforms to experimental evidence for wave nucleation at the grain boundary and to theoretical evidence for drastic difference in the activities of strongly and weakly bound $\underline{N}$ towards $\underline{NH}$ formation. Therefore, none of the external means of the reaction rate control such as reversible surface reconstruction, reactant diffusion, or coverage dependent step constants is necessary for successful modelling the oscillations. The Scheme 1 is based upon common regularities attributing the large number of noble metal surfaces and may be universal. The replacement of Pt(100) plane by the other substrate just needs the step constants correction. It mainly concerns the $\underline{N}$-to-surface bond strength which increases in the row Pt, Rh, Ir; according to Fig. 8, such a correction will decrease the efficiency of reaction (4) thus suppressing $NH_3$ production in the same sequence as it actually takes place [30].

Mathematical modelling has revealed three stable kinetic region-attractors of the steady state, sustained oscillations, and complete reaction inhibition by the adsorbed oxygen atoms. The oscillatory features in Fig. 13 are in good agreement with both experimental data and qualitative consideration of oscillatory cycle in the previous section. Indeed, the total reaction rate ($N_2$ formation) is proportional to $\underline{NO}$, $\underline{N}$ and $\underline{NH}$ coverage and in counter phase with $\underline{N}^{\#}$ and $\underline{O}$ coverage. The $\underline{N}^{\#}$ formation follows the maximum in $\underline{N}$ coverage. The residual $\underline{NH}$ coverage in a middle of oscillation cycle provides the $\varphi$ drop and thus the brightness and sharpness of travelling wave in Fig. 1 [31].

The second principle of thermodynamics truly enables reaction by the Gibbs energy $\Delta G = \Delta H-T\Delta S < 0$, not the enthalpy decrease. This wholly concerns the key step (12) in Sch. (1), where the entropy change under adsorption and conversions in the adsorbed layer can be of importance. Omitting small vibrational components [111], the entropy change $\Delta S$ for a non-localized $\underline{NH}$ state is

$$\Delta S = 2[{}_3S - {}_2S - S_{rot} - S_{compr}],$$

where $_3S -_2S$ is a minimal change of the translational entropy for the *3D-2D* transition under adsorption; $S_{rot} = R\ln(\Sigma_{N_2,H_2} / \Sigma_{NH}) = R\ln 2$ is a symmetry change; $S_{compr} = R\ln(P_{ads} / P_{gas})$ is a compression term normalizing intermolecular distances in the gas phase and in the adsorbed layer. The entropy change for localized *NH* state is

$$\Delta S = 2[_3S - S_{conf} - S_{compr}],$$

where $S_{conf} = R[x\ln x - (x-1)\ln(x-1)]$ is configurational entropy for the *NH* layer at $\theta_{NH}$ = 0.5 ML and then x = $1/\theta$ = 2.

Substitution numerical values gives $\Delta S$ = 15.8 and 137.8 J/mol K, or 4.7 and 41.1 kJ/mol at 298 K per step (12) in Sch. (1), for a non-localized and localized *NH* state, respectively. The former case seems more probable considering high *NH* mobility in Table 6. Both cases result in lifting all curves in Fig. 11 by 4 – 40 kJ/mol and do not affect the conclusions [36].

Traveling waves are often observed during oscillatory heterogeneous catalytic reactions. Meanwhile both uniform distributions of the adsorbed species in a point model and omitting the diffusion steps in a system (10) preclude the spatiotemporal phenomena. This conflict is probably a side effect of desired modelling simplicity since there is no doubt about common origin of the reaction rate and spatiotemporal oscillations.

### 3.5. Other oscillatory reactions

Scheme 1 is quite acceptable for oscillations in the $NH_3$+NO reaction on noble metals [61] under the only condition of ready $NH_{3,ad}$ dissociation. A similar approach can be useful for understanding the spatiotemporal phenomena during $H_2$+$O_2$ reaction on a Pt tip, where the area close to (100) plane is the most active [112, 113]. MIB calculations have been performed for the following reactions:

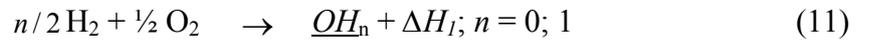

$$n/2\,H_2 + \tfrac{1}{2}\,O_2 \quad \rightarrow \quad \underline{OH}_n + \Delta H_1; \; n = 0;\, 1 \qquad (11)$$

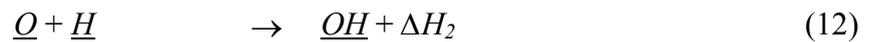

$$\underline{O} + \underline{H} \qquad \rightarrow \quad \underline{OH} + \Delta H_2 \qquad (12)$$

Table 13 shows the results of calculations using additional bond parameters $E_{PtO}$ = 422.6, $E_{OH}$ = 585.3 and $\Delta_O$ = 313.8 kJ/mol determined from the enthalpies of bulk oxides and free $H_2O$ formation. The heat of $O_2$ adsorption on the Pt(111) single crystal 105-120 kJ/mol [106] is inside the range of 80-135 kJ/mol for *OPt₂* and *OPt₃* formation enthalpy in Table 13 that validates the bond parameters.

**Table 13** Thermodynamic properties of the adsorbed *(H)OPt_n* species and enthalpies of reactions (11, 12) on Pt(100) and Pt(111) planes at coverage 1ML (kJ/mol)

| Species | $\nu_{OM}$ | $\nu_{OH}$ | $\Delta H_I$ Pt(100)/Pt(111) | $\Delta H_2$ |
|---------|------------|------------|------------------------------|--------------|
| $H_2O_{gas}$ | - | 0.7886 | -241.8 [43] | - |
| $OPt_4/OPt_3$ | 0.3374 | - | -166.1/-134.7 | - |
| $HOPt_4/HOPt_3$ | 0.2027 | 0.7826 | -242.4/-225.6 | -42.8 |
| $OPt_2$ | 0.5404 | - | -83.3/-78.2 | - |
| $HOPt_2$ | 0.3107 | 0.8334 | -213.7/-207.0 | -96.9 |
| $OPt_1$ | 0.7728 | - | 11.4/22.2 | - |
| $HOPt_1$ | 0.4234 | 0.8865 | -183.9/-176.5 | -161.8 |

Reaction (12) results in stable *OH* species with the larger O-H bond strength ($\nu_{OH}$) than in isolated $H_2O$ molecule; similar to *NH* species, the exothermicity of *O* hydrogenation is larger for the lower number of *O* bonding to surface atoms in Table 13.

Scheme 2 includes the most exothermic steps, according to calculations. Inevitability of *OH* species for $H_2O$ production justifies step (4) in Sch. 2. Intermediate formation of *OH* particles has been revealed during *O*+$H_2$ titration on the Pt(111) single crystal at 105 K [114].

**Scheme 2** The trigger model of oscillations in the $H_2$+$O_2$ reaction on a Pt(100) plane and the calculated step heats $Q_i$ (kJ/mol)

| Step | | | $Q_i$ | |
|------|---|---|-------|---|
| 1. $H_2 + 2*$ | $\leftrightarrow$ | $2\underline{H}$ | 67.0 | |
| 2. ½$O_2$ + * | $\rightarrow$ | $\underline{OPt_4}$ | 166.1 | |
| | $\rightarrow$ | $\underline{OPt_2}$ | 83.3 | |
| 3. $\underline{OPt_4}$ | $\leftrightarrow$ | $\underline{OPt_2}$ | − 82.8 | |
| 4. $\underline{OPt_2}$ + $\underline{H}$ | $\rightarrow$ | $\underline{HOPt_1}$ + * | 67.2 | |
| 5. $\underline{HOPt_1}$ + $\underline{H}$ | $\rightarrow$ | $H_2O$ + * | 24.5 | *Basic reaction route* |
| 6. $2\underline{HOPt_1}$ | $\rightarrow$ | $\underline{OPt_4}$ + $H_2O$ + * | 40.1 | *Trigger step* |

Sustained traveling waves in the $H_2$+$O_2$ reaction on Pt tip can be attributed to the weakly bound, but active in hydrogenation *OPt_2* state similar to $\underline{N}^{\#}$ state in Sch. 1, while intermediate *HOPt_1* species conforms with the preferable valence of oxygen atom. The *OH* combination in a trigger step (6) is more exothermic than basic step (5) in Sch. 2. The wave nucleation starts at critical *OPt_4* coverage enabling the formation of weakly bound *OPt_2* state and a sufficient time term for the further *OH* combination.

A realistic proposition for the CO+O$_2$ reaction on the supported Rh runs "the multiplicity of the steady state is to be expected if thermodynamics permits the existence of a metal and oxide phase on catalyst surface contacting the reactants" [115]. This statement was confirmed later on in a study of the same reaction on the Pt(110) and Pd(100) single crystals [116] and in the reaction of light hydrocarbons oxidation on a Pd foil [117]. Note that the role of different phases consists in providing specific sites to form the adsorbed species of different activities. Assuming common origin of oscillations, the present results are in line with hypothesis [19, 118] modified in the following way. Isothermal oscillations in open heterogeneous catalytic reaction systems are expected under the multiplicity of reaction intermediates, fairly different in activity, providing the steady state and the reaction rout multiplicity. The surface structure, subsurface oxygen, and the oxide phase are generally responsible for inactive adsorbed state, while the proper structure and unoxidized surface provide the active state [19, 30]. Vacuum conditions disable the bulk phase formation. The function of active and inactive phase is thus fulfilled by $\underline{N}^{\#}$ and $\underline{N}$ states or "surface adsorbate phases" [119] different in activity towards the $\underline{NH}$ species formation as a key step of the temporary reaction channel.

# Conclusions

Unique catalytic properties of the transition and noble metals encouraged a great number of fundamental and applied studies. This manuscript highlighted the general regularities in the field on the grounds of strong interconnection between catalytic, kinetic and thermodynamic behaviour of a heterogeneous catalytic system. The trials of the catalytic $NH_3$ synthesis and the oscillatory $NO+H_2$ reactions have revealed that the thermodynamics of the local structure determines the properties and multiplicity of the reaction intermediates enabling the peculiar macroscopic kinetics and specific catalytic activity.

A realistic model of quantitative correlation between the local structure and the activity of catalytic centers is proposed. The model evaluates a strong advantage of surface defects in $NH_{ad}$ species formation, and therefore in surface waves nucleation as compared to perfect terraces. With respect to ammonia synthesis, the model has specified the resonant catalytic centers on metal surfaces in close agreement with experimental findings. The basal planes of the noble metals are less active as compared to Fe- and Ru-based catalysts, whereas small Pt, Ir and Rh clusters are expected to reveal an extraordinary catalytic activity.

Isothermal rate oscillations in open heterogeneous catalytic reaction systems are expected under the multiplicity of key reaction intermediates, fairly different in activity, providing the steady state and reaction rout multiplicity. A switching between the active and inactive kinetic brunches gives rise to the explosive coverage changeover that can be visualized as a traveling wave. A single pattern of oscillations in the $NO+H_2$ reaction includes the key role of $NH_{ad}$ species, which follow the "easy-come-easy-go" principle and provide the catalytic removal of strongly bound adsorbed nitrogen. Mathematical modeling at fixed step constants revealed three kinetic region-attractors of the steady state, regular oscillations, and complete reaction inhibition by the adsorbed oxygen. The driving forces, the feedback mechanism and chemical interactions within the traveling waves are discussed in detail and can be clearly understood.